\documentclass[aps,pre,nofootinbib,nobibnotes,floatfix,tightenlines,twocolumn, superscriptaddress,notitlepage]{revtex4-1}

\usepackage{epsfig, dsfont, amssymb, amsmath, amsthm, amsfonts, amsbsy, mathrsfs, bm, tabularx, graphicx} 
\usepackage{multirow,float,setspace,ragged2e,upgreek, tikz, mathtools, siunitx}
\usepackage{color, enumitem, hyperref, cleveref, caption, gensymb}
\usepackage{lipsum}

\hypersetup{hidelinks}

\DeclareCaptionJustification{justified}{\justifying}

\usepackage[a4paper, total={7in, 10in}]{geometry}

\newcommand{\STable}{Supplementary Table }
\newcommand{\SFig}{Supplementary Fig.\ }

\newcommand{\SVid}{Supplementary Movie }

\begin{document}

\title{Predicting mosquito flight behavior using Bayesian dynamical systems learning}

\author{Christopher Zuo}
\thanks{These authors contributed equally.}
\affiliation{George W. Woodruff School of Mechanical Engineering, Georgia Institute of Technology, Atlanta, GA 30332, USA}

\author{Chenyi Fei}%
\thanks{These authors contributed equally.}
\affiliation{Department of Mathematics, Massachusetts Institute of Technology, Cambridge, MA 02139, USA}

\author{Alexander E. Cohen}
\thanks{These authors contributed equally.}
\affiliation{Department of Mathematics, Massachusetts Institute of Technology, Cambridge, MA 02139, USA}

\author{Soohwan Kim}
\affiliation{George W. Woodruff School of Mechanical Engineering, Georgia Institute of Technology, Atlanta, GA 30332, USA}
\author{Ring T. Carde}
\affiliation{Department of Entomology, University of California, Riverside, CA 92521, USA}
\author{Jörn Dunkel}
\email[Corresponding author:]{dunkel@mit.edu}
\affiliation{Department of Mathematics, Massachusetts Institute of Technology, Cambridge, MA 02139, USA}
\author{David L. Hu}
 \email[Corresponding author:]{hu@me.gatech.edu}
\affiliation{George W. Woodruff School of Mechanical Engineering, Georgia Institute of Technology, Atlanta, GA 30332, USA}
\affiliation{School of Biological Sciences, Georgia Institute of Technology, Atlanta, GA 30332, USA}

\begin{abstract}
Mosquito-borne diseases cause several hundred thousand deaths every year. Deciphering mosquito host-seeking behavior is essential to prevent disease transmission through mosquito capture and surveillance.  Despite recent substantial progress, we currently lack a comprehensive quantitative understanding of how visual and other sensory cues guide mosquitoes to their targets.  Here, we combined 3D infrared tracking of \textit{Aedes aegypti} mosquitoes with Bayesian dynamical systems inference to learn a quantitative biophysical model of mosquito host-seeking behavior.  Trained on more than 20,000,000 data points from mosquito free-flight trajectories recorded in the presence of visual and carbon dioxide cues, the model accurately predicts how mosquitoes respond to human targets. Our results provide a quantitative foundation for optimizing mosquito capture and control strategies, a key step towards mitigating the impact of mosquito-borne diseases. 
\end{abstract}

\maketitle

% figurename
\renewcommand{\thesection}{\Roman{section}}
\renewcommand{\thesubsection}{\Alph{subsection}}
\renewcommand{\thetable}{\arabic{table}}
\renewcommand{\thefigure}{\arabic{figure}}
\renewcommand{\figurename}{Fig.}
\setcounter{figure}{0}
\setcounter{equation}{0}
\setcounter{table}{0}
\setcounter{section}{0}

\subsection*{Introduction}

Mosquitoes are often called the world's most dangerous animals as they transmit diseases such as malaria, dengue virus, and Zika, which collectively cause over 770,000 deaths annually~\cite{omodior2018mosquito,wang2023novel}. 
Among the 3500 mosquito species, approximately 100 have evolved to be anthropophilic, meaning they preferentially target human hosts~\cite{mcbride2014evolution,yee2022robust}. {\it Aedes~aegypti} are one such anthropophilic mosquito species, and females employ a suite of cues to locate a human host~\cite{hinze2022chapter,van2015mosquitoes,de2020neuro,cribellier2018flight,liu2019general,warrant2011vision,alonso2022olfactory,van2015mosquitoes,gillies1981field,gillies1974evidence}. These cues have mostly been tested in wind tunnels, where winds carry odors like carbon dioxide downwind \cite{sumner2022primacy,cooperband2006comparison,van2015mosquitoes}.  Up to distances of 10 meters downwind, emitted carbon dioxide both triggers mosquito flight towards the host and lowers their threshold of detection of skin odors. In addition to chemical cues, visual stimuli can guide mosquito flight to within meters of the host \cite{van2015mosquitoes}. Once in direct contact with the host, cues like skin odor, heat, and humidity\cite{laursen2023humidity} aid the mosquito in landing and choosing a spot for probing for capillaries~\cite{choumet2012visualizing}. 
Despite decades of study, how mosquitoes integrate these signals to locate hosts remains poorly understood. Due to our inability to predict mosquito flight behavior, commonly used devices such as suction traps are only 10-50 percent effective in capturing incoming mosquitoes~\cite{amos2020attraction,cooperband2006orientation,cooperband2006comparison}.  An improved understanding of mosquito flight and host-seeking behaviors can help inform the design of more efficient intervention strategies for insect-borne diseases as well as the development of mosquito-resistant infrastructure in private and public spaces~\cite{jatta2021impact}.

Previous investigations of mosquito host-seeking behavior were typically limited to small numbers of tethered mosquitoes or mosquitoes flying in wind tunnels~\cite{van2015mosquitoes,majeed2017detection,cooperband2006comparison,cooperband2006orientation}. 
These studies often report statistics of various intuitive metrics, such as the number and positions of mosquito landings, rather than providing continuous flight information~\cite{de2007selection,dekker1998selection,hawkes2016seeing,sinhuber2019three,spitzen20133d}.
However, flight trajectories are crucial for understanding mosquito host-seeking behavior, as mosquitoes use the time-integrated response of their sensory information to make a host-seeking decision.\cite{alonso2022olfactory} Moreover, mosquito eyes limit the resolution and distance at which they can detect visual stimuli~\cite{muir1992aedes,muir1992vision,hawkes2022chapter}.
Capturing the 3D sensory information and the 3D flight trajectories necessitates a data-driven approach due to the size and dimensionality of the resulting datasets.
The data-driven identification of interpretable biophysical models that accurately predict 3D mosquito flight in the presence of 3D sensory stimuli under realistic environmental conditions will not only substantially enhance our understanding of mosquito host-seeking behavior, but can also provide guidance for data-driven model inference for other species, such as bees, ants, starlings, and humans \cite{mlot2011fire,peleg2018collective,berman2014mapping,mendez2018density,attanasi2014information,bozek2021markerless}. Here, we introduce such framework by combining 3D inference tracking experiments with Bayesian dynamical systems inference~\cite{tipping2001sparse, wipf2004sparse, pan2015sparse, yuan2019data, fuentes2021equation, huang2022sparse, hirsh2022sparsifying,mangan2016inferring,bruckner2020inferring}.   
\par
To establish a foundational database of various mosquito flight behaviors, we perform 3D tracking experiments on female {\it Aedes~aegypti} mosquitoes interacting with visual and CO$_2$ cues, both individually and in combination~\cite{keller2020optical,patt2024optical}. Since {\it Aedes} thrive in urban areas, our experiments are performed in relatively windless conditions, as typical of human dwellings.  Across 20 experiments, we generate 53,669,795 data-points and over 477,220 mosquito trajectories, exceeding by orders of magnitude previous attempts to quantitatively measure mosquito behavior (\STable\ref{tabS:trajcompare} and \STable\ref{tabS:trajtotal}) ~\cite{kennedy1940visual,geier1999influence,spitzen20133d,hawkes2016seeing,cribellier2018flight,sinhuber2019three,cribellier2020lure,amos2020attraction,amos2020attraction2,hinze2021mosquito}.  

We then apply sparse Bayesian dynamical systems inference~\cite{tipping2001sparse, wipf2004sparse, pan2015sparse, yuan2019data, fuentes2021equation, huang2022sparse, hirsh2022sparsifying,mangan2016inferring,bruckner2020inferring} to learn quantitative models of mosquito behavior directly from the mosquito flight trajectories.
This data-driven method produces a predictive dynamical model that does not require human-defined behavioral labels, yet allows us to compute various behavioral traits of interest.

\subsection*{Results}

\subsubsection*{Tracking of 3D mosquito trajectories}

\begin{figure*}[!t]
    \centering
    \includegraphics[width = .6\textwidth]{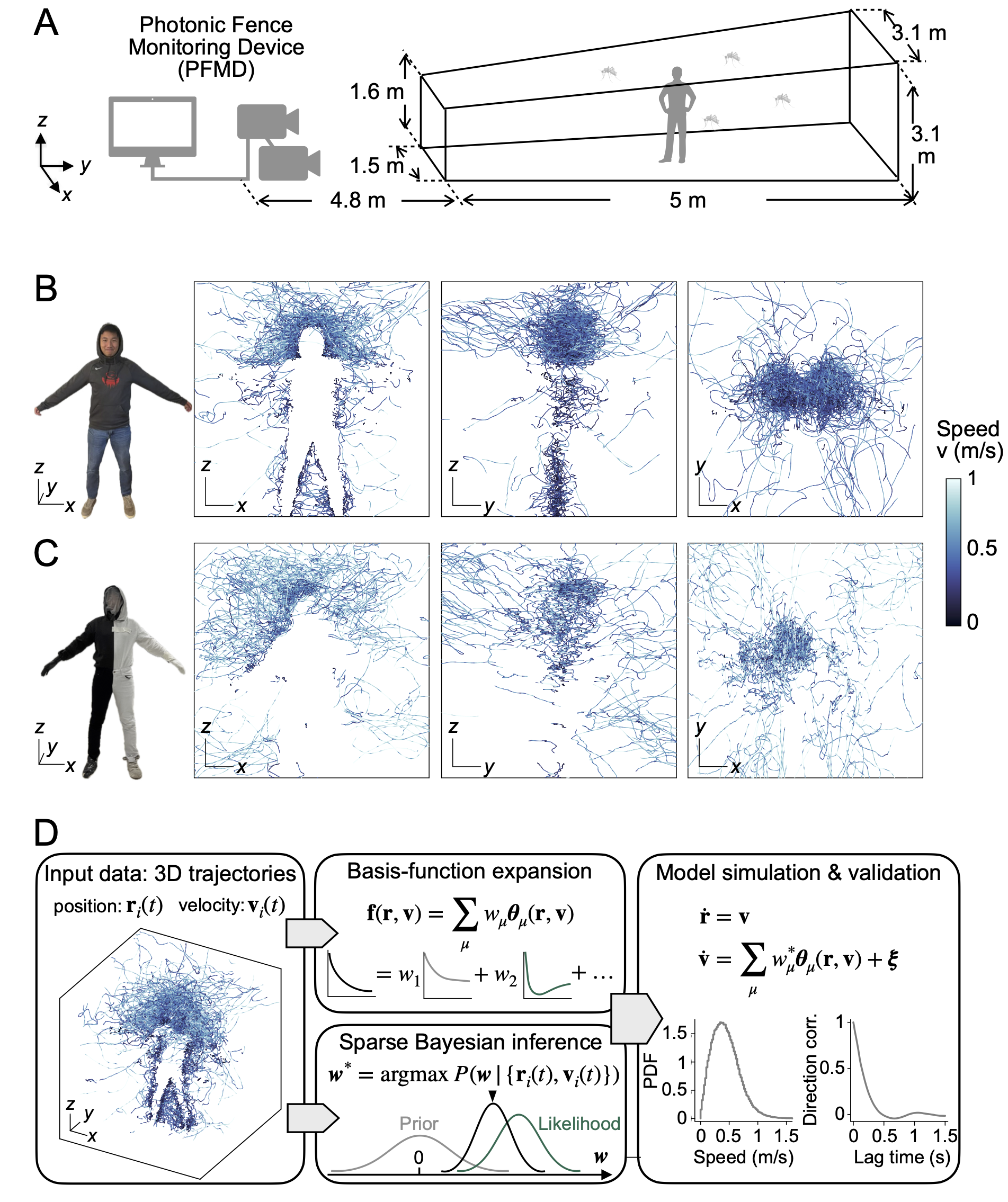}
    \captionsetup{labelsep = period, justification=justified,labelfont=bf,singlelinecheck = false}
    \caption[3D tracking of individual mosquitos enables dynamical inference of mosquito flight behaviors.]
    {
    {3D tracking of individual mosquitos enables dynamical inference of mosquito flight behaviors.} 
    ({\bf A}) Schematics of the experimental setup.
    ({\bf B}) 3D views of mosquito trajectories around a standing human subject wearing a normal outfit ({\it left}), colored by flight speed.
    ({\bf C}) 3D views of mosquito trajectories around a standing human subject wearing a half-black, half-white outfit ({\it left}), colored by flight speed.
    ({\bf D}) Inference of dynamical models for mosquito flight behaviors. {\it Left}: Inputs are 3D time-series data for mosquito positions $\mathbf{r}(t)$ and velocities $\mathbf{v}(t)$. {\it Middle}: The unknown behavioral force $\mathbf{f}$ is decomposed on a set of basis functions $\boldsymbol{\theta}_\mu$, where the decomposition coefficients $w_\mu$ are learned from input data using sparse Bayesian inference. {\it Right}: The learned dynamical model can be simulated to generate synthetic time-series data for validation against input data, comparing them based on ensemble statistics such as speed distribution and directional correlation function, which are not used in model inference. Scale bars labeled with x, y, and z: 25 cm. For visual clarity, 50 \% of trajectories are shown.
  }
    \label{fig:demo}
\end{figure*}

We conducted the experiments at 28\textdegree C and 45\% humidity inside a trapezoidal mesh enclosure of 5 m depth shown in Fig.~\ref{fig:demo}A. Facing the enclosure is the photonic fence monitoring device (PFMD) which is a set of dual infrared cameras surrounded by infrared 960 nm LEDs. The LEDs illuminate a scene against a retro-reflective background allowing for the dual cameras to obtain stereoscopic images of insect positions in space at 0.01-s time resolution.  Mosquitoes are released at the front of the chamber and various targets (human, styrofoam spheres, carbon dioxide sources) are placed 8 m away from the camera.

We first apply the PFMD to image mosquito trajectories around a human subject wearing dark clothing (Fig.~\ref{fig:demo} A \& B). In this experiment, our system tracked 50 mosquitoes for 20 minutes and recorded on average 22,000 positional trajectories of mosquitoes in 3D, generating on average 1,961,000 data points. A trajectory represents a segment of the full flight path of one mosquito. Our experiment shows {\it Ae.~aegypti} mosquitoes primarily target the human head (Fig.~\ref{fig:demo}B), consistent with previous studies of bite location \cite{de2007selection}. As shown by the darker colors of the trajectories in Fig.~\ref{fig:demo}B, mosquitoes decelerate near the human head and body, suggesting preparation for landing.  The complex 3D tracks reflect the mosquitoes' response to the multiple cues detected. 

To demonstrate the importance of visual cues in windless situations, we tracked mosquitoes around a human subject wearing a Janus outfit with the left side white and the right side black.  As can be seen from the front and top views, the trajectories are primarily concentrated on the black side, despite other cues such as carbon dioxide and odor remaining symmetric (Fig.~\ref{fig:demo}C, \SVid{1}, and \SVid{2}).  Due to the complexity of evaluating the cues around a human subject, we pivot from human experiments and proceed with experiments using simple objects that present individual cues. These controlled experiments allow for simple formulation of free-flight characteristics and a minimal model of {\it Ae.~aegypti} in response to individual cues.  At the end of our study, we revisit the prediction of mosquito behavior around a human subject using our minimal model. 

\subsubsection*{Data-driven inference facilitates the discovery of dynamical models for mosquito flight}

To model mosquito flight behavior, we consider a stochastic Langevin dynamics
\begin{equation}
    \frac{\mathrm{d}\mathbf{v}}{\mathrm{d}t}  = \mathbf{f}(\mathbf{r}, \mathbf{v}) + \boldsymbol{\xi} \label{eqn:langevin}
\end{equation}
where $\mathbf{f}$ is a deterministic force (per mosquito body mass) that depends on the spatial position $\mathbf{r}$ of a mosquito and its flight velocity $\mathbf{v}$, and $\boldsymbol{\xi}$ is a noise term that accounts for fluctuations. Here, we model $\boldsymbol{\xi}$ as Gaussian white noise that satisfies $\langle \boldsymbol{\xi}(t) \boldsymbol{\xi}(t') \rangle = \Delta \delta(t-t^\prime)$, where $\Delta$ characterizes the magnitude of noise. To infer mosquito flight behavior from data, we approximate the unknown forces $\mathbf{f}(\mathbf{r}, \mathbf{v})$ using a basis-function expansion \cite{frishman2020learning,ronceray2024learning,bruckner2020inferring,bruckner2021learning} (Fig.~\ref{fig:demo}D)
\begin{gather}
\mathbf{f}(\mathbf{r}, \mathbf{v}) = \sum_\mu w_\mu \boldsymbol{\theta}_\mu(\mathbf{r},\mathbf{v}), \label{eqn:basis-function-expansion}
\end{gather}
where $\boldsymbol{\theta}_\mu$ represents orthogonal polynomials with weight functions (see Supplementary Materials), and the coefficients $w_\mu$ encode all the information about mosquito behavior. This linear expansion in Eq.~(\ref{eqn:basis-function-expansion}) simplifies the behavioral inference problem by reducing it to a linear regression problem.

To learn $w_\mu$ from data, we use a Bayesian approach to find the most probable coefficients $w_\mu^*$ that maximize the posterior probability $P(\{w_\mu\} | \{\mathbf{r}_i(t),\mathbf{v}_i(t)\})$ of $w_\mu$ given the experimental trajectories $\{\mathbf{r}_i(t),\mathbf{v}_i(t)\}$. In Bayesian statistics, the posterior is proportional to the product of two components (Fig.~\ref{fig:demo}C): a likelihood function $P(\{\mathbf{r}_i(t),\mathbf{v}_i(t)\} | \{w_\mu\})$ of the observed trajectories given the coefficients $\{w_\mu\}$, and a prior probability $P(\{w_\mu\})$ that reflects our preference for the desired coefficients $\{w_\mu\}$. Assuming that each mosquito track is independent and neglecting mosquito-mosquito interactions, which we find to be minor (see \SFig\ref{figS:pair_traj}), Eq.~(\ref{eqn:basis-function-expansion}) yields a multivariate Gaussian likelihood function (see Supplementary Materials). To prevent overfitting to noisy data, we impose sparsity on $\{w_\mu\}$, favoring a parsimonious representation of smooth functions $\mathbf{f}(\mathbf{r},\mathbf{v})$ with few nonzero $w_\mu$. Following previous works on sparse Bayesian inference \cite{tipping2001sparse}, we use a sparsity-promoting Gaussian prior $P(w_\mu) \propto \exp( - \frac{w_\mu^2}{2\sigma^2_\mu})$
where the variance hyperparameters $\sigma^2_\mu$ control the degree of sparsity in $\{w_\mu\}$. To estimate the coefficients $w_\mu^*$, the noise magnitude $\Delta^*$, and the hyperparameters $\sigma^2_\mu$, we use an expectation-maximization algorithm to iteratively maximize the posterior and refine model parameters (see Supplementary Materials).

To explore a series of candidate models with varying degrees of sparsity, we apply sequential thresholding to $w_\mu^*$ by progressively eliminating coefficients with small absolute values \cite{brunton2016discovering,mangan2017model,mangan2019model}. Although including more terms in the expansion in Eq.~(\ref{eqn:basis-function-expansion}) can generally improve the fit to data, overly complex models may capture noise in the data rather than the true underlying forces $\mathbf{f}(\mathbf{r}, \mathbf{v})$, and exhibit poor predictive power. Thus, to identify the model with optimal sparsity, we use the Bayesian information criterion \cite{neath2012bayesian} to select the model that best explains the data with the least number of terms in Eq.~(\ref{eqn:basis-function-expansion}) (see ~\SFig\ref{figS:levitation}). 
We validated our model identification pipeline on synthetic data of mosquito flight trajectories (\SFig\ref{figS:demo_velpot} and \SFig\ref{figS:demo_response}). Below, we apply this pipeline to experimental data to uncover dynamical models of mosquito flight behavior.

\subsubsection*{Mosquitoes exhibit two distinct modes of free flight}
% The findings from the experiments revealed key insights into mosquito behavior. 

\begin{figure*}[!t]
    \centering
    \includegraphics[width = .72\textwidth]{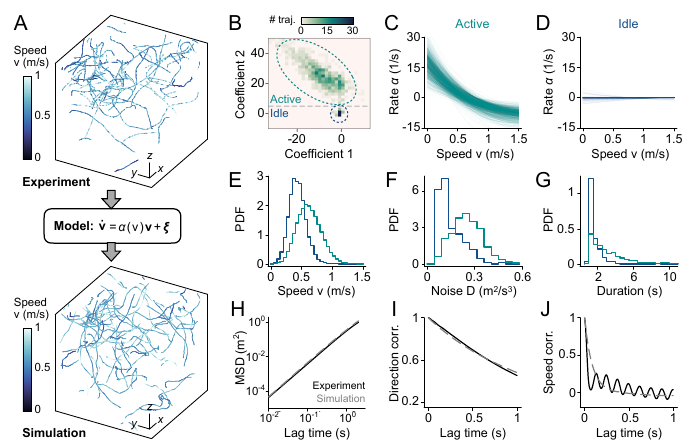}
    \captionsetup{labelsep = period, justification=justified,labelfont=bf,singlelinecheck = false}
    \caption[Free flying mosquitos exhibits two distinct behaviors.]
    {
    {Free flying mosquitos exhibits two distinct behaviors.} 
    ({\bf A}) ({\it Top}) 
 Representative mosquito trajectories from experiments and ({\it Bottom}) simulations of the learned model. Scale bars labeled with x, y, and z: 25 cm. For visual clarity, only 10 \% of total trajectories are shown.
    ({\bf B}) 2D histogram of learned decomposition coefficients shows two groups of trajectories, an active group (green) and an idle group (blue).
    ({\bf C}, {\bf D}) The inferred rates $\alpha$ at varying speed $v$ for ({\bf C})~the active group and ({\bf D}) the idle group.
    ({\bf E} -- {\bf G}) Probability distributions of ({\bf E}) the measured flight speed, ({\bf F}) the inferred velocity diffusion coefficient, and ({\bf G}) the duration of the trajectories for the two groups.
    ({\bf H}) Mean squared displacement, ({\bf I}) directional correlation, and ({\bf J}) speed correlation are compared between experiments (black) and simulations of the learned model (gray).
  }
    \label{fig:free}
\end{figure*}

We first demonstrate our model inference and selection framework on experimental trajectories of free-flight mosquitoes in the absence of sensory stimuli (Fig.~\ref{fig:free}). We film 50 mosquitoes for 20 minutes in an empty room.  To model the flight behavior, we consider three forces acting on a mosquito: a thrust in the direction of flight $\hat{\mathbf{v}}$, a constant gravitational force, and a levitational force in $\hat{\mathbf{z}}$ that counteracts gravity. Our inference results indicate that the best model includes a constant levitation force that precisely balances the gravitational force (\SFig\ref{figS:levitation}), suggesting minimal preference for the $\pm\hat{\mathbf{z}}$ direction of flight. Therefore, we ignore these two forces in subsequent analysis. This simplifies the free-flight Langevin equation to $\frac{\mathrm{d}\mathbf{v}}{\mathrm{d}t} = \alpha(v)\mathbf{v} + \boldsymbol{\xi}$ (Fig.~\ref{fig:free}A), where the learned thrust $\alpha(v) \mathbf{v}$ corresponds to a force derived from a speed potential $U(v)$ with $\alpha(v) = -U'(v)/v$.

To further examine behavioral variability, we performed inference on individual tracks and projected the thrust factor $\alpha(v)$ onto the two most relevant basis functions. The learned coefficients reveal two distinct clusters, one corresponding to an active state of flight and the other to an idle state (Fig.~\ref{fig:free}B).
The active state tends to maintain a constant flight speed of approximately 0.7 m/s, with the mosquito accelerating (positive $\alpha$) if its flight speed is too low and decelerating (negative $\alpha$) if its speed is too high (Fig.~\ref{fig:free}C). This flight speed is consistent with experimental values (0.4-1.8 m/s) measured for other mosquito species \cite{gillies1981field}.  
The idle state corresponds to a lack of baseline speed with no accelerations or decelerations (zero $\alpha(v)$) (Fig.~\ref{fig:free}D), allowing noise to dictate the dynamics of flight velocity. 
The active state exhibits higher mean speed (Fig.~\ref{fig:free}E), stronger inferred noise (Fig.~\ref{fig:free}F), and longer trajectory duration (Fig.~\ref{fig:free}G) than the idle state, suggesting that the active state may reflect exploration behavior while the idle state corresponds to preparation for landing. Indeed, the idle state is more often observed in trajectories near the ceiling of the chamber (\SFig\ref{figS:zcdf_active_idle}).

To validate the learned model, we simulate the inferred dynamical equations with the same initial conditions as the experimental data. The simulated trajectories (Fig.~\ref{fig:free}A; bottom) qualitatively resemble the observed trajectories (Fig.~\ref{fig:free}A; top). To further quantify the similarity, we compute three time-lagged statistics, $\langle S(t, t+\tau)\rangle$, which compare two time points along a trajectory separated by a lag time $\tau$ (Figs.~\ref{fig:free}H--J): the mean squared displacement (MSD), defined by $S_{\mathbf{r}}(t, t+\tau)=|\mathbf{r}(t) - \mathbf{r}(t+\tau)|^2$, the directional correlation, defined by $S_{\hat{\mathbf{v}}}(t, t+\tau)=\hat{\mathbf{v}}(t) \cdot \hat{\mathbf{v}}(t+\tau)$, and the speed correlation,  defined by $S_v(t, t+\tau)=(v(t)-\langle v\rangle)(v(t+\tau)-\langle v\rangle)$.
Here, $\langle \cdot \rangle$ denotes the average over all time points $t$ of all trajectories.
The MSDs of both simulated and observed trajectories show a ballistic flight at short lag times, followed by a transition to more diffusive movement at around $\tau = 1~\mathrm{s}$ (Fig.~\ref{fig:free}H). However, the experimental trajectories are not long enough to fully capture the diffusive behavior beyond this timescale. The directional correlation of mosquito flight roughly follows an exponential decay $e^{-\tau/\tau_\mathrm{d}}$ (Fig.~\ref{fig:free}I), representing straight flight over short time intervals before making turns after $\tau_\mathrm{d}\approx 1.3$ s. The speed correlation decreases toward zero with the lag time, a characteristic of the Langevin equation. Interestingly, the experimental speed correlation exhibits oscillations with a period of roughly 0.125~s (Fig.~\ref{fig:free}J).
We speculate the oscillation could reflect sensory-feedback control, as found for fruit flies \cite{dickinson2016aerodynamics,ristroph2013active}, as well as oscillations in lift due to upstroke and downstroke \cite{liu2020aerodynamic}. 
The close agreement between the experimental data and the model simulations demonstrates that our inference framework effectively learns the Langevin equation and behavioral forces governing mosquito flight dynamics in the absence of sensory cues.

\subsubsection*{Mosquitoes are attracted to visual cues and \texorpdfstring{CO$_2$}{CO2} plumes through differential responses}

\begin{figure*}[!t]
    \centering
    \includegraphics[width = 0.96\textwidth]{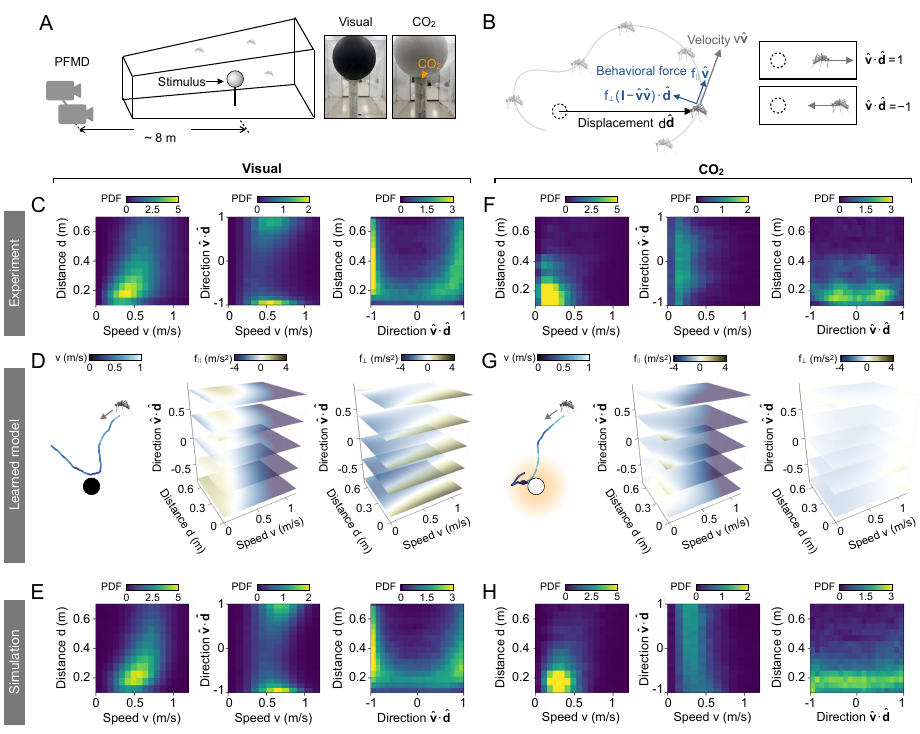}
    \captionsetup{labelsep = period, justification=justified,labelfont=bf,singlelinecheck = false}
    \caption[Dynamical inference of mosquito flight behaviors reveals different responses to stimuli.]
    {
    {Dynamical inference of mosquito flight behaviors reveals different responses to stimuli.} 
    ({\bf A}) Schematics of the experimental setup.
    ({\bf B}) Illustration of a mosquito flying with a velocity $v\hat{\bf v}$ and a displacement $d\hat{\bf d}$ relative to the stimulus (dashed circle). The behavioral forces are decomposed into two directions, one along ${\bf v}$ ($f_\parallel$) and one perpendicular to ${\bf v}$ $(f_\perp)$, and are expressed as functions of three scalar variables: the flight speed $v$, the distance $d$ to the stimulus, and the flight direction $\hat{\bf v} \cdot \hat{\bf d}$.  $\hat{\bf v} \cdot \hat{\bf d} = -1$ and $1$ indicate mosquito flight toward and away from the stimulus, respectively.
    ({\bf C} -- {\bf E}) Mosquito response to the visual stimulus. ({\bf C}) 2D density maps of recorded mosquito trajectories reveal mosquito attraction to the visual stimulus with bidirectional flights either toward or away from the stimulus. ({\bf D}) {\it Left}: A typical 2D simulation trajectory of the learned model of mosquito response to visual cues (black circle). Heatmaps of the inferred mosquito behavioral forces ({\it Middle}) $f_\parallel$ and ({\it Right}) $f_\perp$ in response to visual stimulus. For $f_\parallel$, yellow and blue represent increase and decrease in speed, respectively. White denotes steady-state speed with $f_\parallel=0$. For $f_\perp$, yellow and blue represent turning away from and toward the stimulus, respectively. ({\bf E}) 2D density maps of simulated mosquito trajectories show quantitative agreement with the experimental data in {\bf C}. 
    ({\bf F} -- {\bf H}) Mosquito response to the CO$_2$ cues. Plots correspond to {\bf C} -- {\bf E}. 
  }
    \label{fig:stimulus}
\end{figure*}

Having established a basis for mosquito behavior, we next add attractive cues into the environment. 
In this work, we focus on visual and CO$_2$ cues (Figs.~\ref{fig:stimulus}--\ref{fig:trajectory}), but our experimental and inference framework should be directly applicable to other stimuli such as odor and heat.
We use an 8-inch ($\approx$0.2 m) diameter sphere mounted on a white pole as the attractive target for mosquitoes. To minimize boundary effects, the sphere is placed at the center of the chamber (Fig.~\ref{fig:stimulus}A). A black sphere is used as a visual target due to its high contrast with the surrounding white walls.
 The CO$_2$ target is a white sphere with piping releasing  CO$_2$ at a rate of 0.24~$\mathrm{L/min}$, comparable to the CO$_2$ emission rate of human breathing \cite{pleil2021physics}. 
 % A black sphere releasing $CO_2$ is used for a combined cue target.

We describe the state of a mosquito by its displacement $\mathbf{d}$ from the spherical target and body velocity $\mathbf{v}$ (Fig.~\ref{fig:stimulus}B).  The dot product $\hat{\mathbf{v}} \cdot \hat{\mathbf{d}}$ represents the orientation of mosquito flight,
with $\hat{\mathbf{v}} \cdot \hat{\mathbf{d}} = -1$ indicating flight directly towards the target (taxis), and $\hat{\mathbf{v}} \cdot \hat{\mathbf{d}} = 1$ indicating flight away from the target. 
 % $f_\parallel$ is analogous the mosquito's throttle, or accelerator pedal: a positive $f_\parallel$ indicates that the mosquito is accelerating in its current direction of motion; negative indicates deceleration. $f_\perp$ indicates the veering of a mosquito towards the stimulus, with negative $f_\perp < 0$ means turning towards the target.  A mosquito that is circling the target has a constant $f_\perp = -v^2/R$ where $R$ is the turning radius. A zero value for $f_\perp$ indicates that there is no acceleration towards or away from the target: i.e., mosquito keeps flying in their same direction.  A positive value for $f_\perp$ indicates acceleration away from the target, as would occur to avoid a crash landing.
To visualize mosquito flight patterns around the target, we present heatmaps of trajectory densities at varying distances $d$ to the target, flight speeds $v$, and flight orientation $\hat{\mathbf{v}} \cdot \hat{\mathbf{d}}$ (Fig.~\ref{fig:stimulus}C and F).

Our data reveal distinct mosquito flight patterns in response to visual or CO$_2$ cues.  We consider the visual cue first (Fig.~\ref{fig:stimulus}C, left). The diffuse yellow region in the density heatmap shows slower mosquitoes concentrated in a zone of radius 0.4 m from the target. As the distance exceeds about 0.4 m, the speed distribution returns to the baseline observed in the absence of a stimulus (Fig.~\ref{fig:free}E), which is consistent with the mosquito's visual range calculated from their eye's minimal resolvable angle of $12.3\degree$ \cite{muir1992aedes}. 
The mosquito’s flight is approximately bidirectional, with trajectory densities concentrated around $\hat{\mathbf{v}} \cdot \hat{\mathbf{d}} = \pm 1$ (Fig.~\ref{fig:stimulus}C, middle and right), indicating movement either directly towards or away from the visual target.  We interpret this behavior as either attraction to the target or rejection of the target, after having already explored it and failing to find a blood meal.
Indeed, a closer examination of trajectories (Fig.~\ref{fig:trajectory}E) shows that although mosquitoes are attracted to the visual target, they often perform a ``fly-by,'' and do not consistently engage with it due to the absence of additional host cues (\SVid{2}). Nevertheless, the visual cue alone can attract a high density of mosquitoes.  To demonstrate that attraction to black spheres can be extended to arbitrary shapes, we wrote ``GT'' in black using a 40-inch font on the white floor, attracting mosquitoes approximately uniformly to each part of the lettering (\SFig\ref{figS:logo}).

When presented with the CO$_2$ cue,  mosquitoes perform ``double-takes'' or tumbling behavior with decreased speed and increased turning that keeps them in the local vicinity of the target (Fig.~\ref{fig:trajectory}H and \SVid{3}).  This tumbling has  no directional preference, with $\hat{\mathbf{v}} \cdot \hat{\mathbf{d}}$ uniformly distributed between -1 and 1 (Fig.~\ref{fig:stimulus}F, middle and right). CO$_2$-induced tumbling is thus a form of kinesis, which is undirected movement, increasing their concentration around the target relative to that of a visual cue (Fig.~\ref{fig:stimulus}F, left).  Mosquitoes can sense CO$_2$ concentrations as low as 103 ppm \cite{grant2007age}. Numerical simulations of our experimental setup show that the zone of detectable CO$_2$ extends to 0.5 m from the source (\SFig\ref{figS:co2}), which aligns with our observation of increased mosquito density within 0.3~m of the target.

To model mosquito response to sensory cues, we decompose the behavioral force, 
\begin{equation}
\mathbf{f} = f_\parallel \hat{\mathbf{v}} + f_\perp (\mathbf{I} - \hat{\mathbf{v}}\hat{\mathbf{v}})\cdot \hat{\mathbf{d}},    
\end{equation}
 into two components (Fig.~\ref{fig:stimulus}B): a longitudinal force $f_\parallel$ along the direction of flight $\hat{\mathbf{v}}$ and a transverse force $f_\perp$. The longitudinal force $f_\parallel$ represents the mosquito's throttle, with positive and negative values corresponding to acceleration and deceleration, respectively, in the mosquito's direction of motion. The transverse force $f_\perp$ represents the mosquito's turning, with negative $f_\perp$ indicating turning towards the target (taxis) and positive $f_\perp$ indicating turning away from it. We expand $f_\perp$ and $f_\parallel$ onto basis functions, given by tensor products of univariate scalar functions of $v$, $d$, $\hat{\mathbf{v}}\cdot\hat{\mathbf{d}}$, and apply Bayesian inference to estimate the force magnitudes. 

Our inference results support different responses to visual and CO$_2$ cues, which we discuss in turn. In response to visual cues, the longitudinal force $f_\parallel$  resembles a speed potential force as in Fig.~\ref{fig:free}C, with the preferred speed, indicated by white heatmap regions where $f_\parallel=0$, gradually decreasing as the mosquitoes approach the target (Fig.~\ref{fig:stimulus}D and \SFig{8}). The transverse force $f_\perp$ reorients mosquitoes toward the visual target when they are far from it but directs them away when they are too close, likely to prevent collisions and facilitate departure if no additional attractants are detected. 

In contrast, in response to CO$_2$ cues (Fig.~\ref{fig:stimulus}G and \SFig\ref{figS:learnedforces}), $f_\perp$ is much weaker than its counterpart for visual cues, showing white regions near $f_\perp = 0$ throughout. The $f_\parallel$ force corresponds to a deceleration from a flight speed of 0.7 m/s to 0.2 m/s, when the distance to the CO$_2$ source drops below 0.4 m. This deceleration alone is sufficient to explain the higher concentration of mosquito tracks near the CO$_2$ stimuli \cite{Tailleur}. Simulations of the learned models quantitatively match the experimental data, including the density distributions around the stimuli (Fig.~\ref{fig:stimulus}E and H), bidirectional flight toward or away from the visual cue (Fig.~\ref{fig:trajectory}D--F and \SVid{4}), and tumbling behavior near the CO$_2$ source (Fig.~\ref{fig:trajectory}G--I and \SVid{5}). Thus, our inference framework provides a powerful tool for discovering quantitative models for mosquito behavior in response to individual cues.

%Combining $CO_2$ and Visual
\subsubsection*{Mosquitoes combine visual and \texorpdfstring{CO$_2$}{CO2} cues to target human}

\begin{figure*}[!t]
    \centering
    \includegraphics[width = \textwidth]{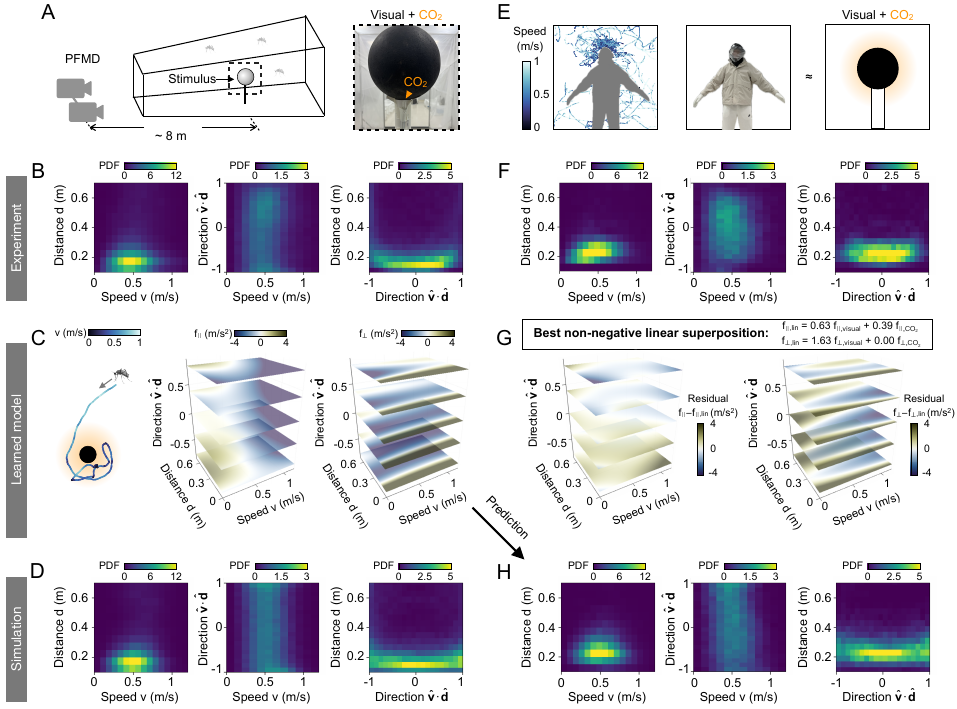}
    \captionsetup{labelsep = period, justification=justified,labelfont=bf,singlelinecheck = false}
    \caption[Mosquitoes combine visual and CO$_2$ cues to target human hosts.]
    {
    {Mosquitoes combine visual and CO$_2$ cues to target human hosts.} 
    ({\bf A}) Schematics of the experimental setup.
    ({\bf B} -- {\bf D}) Mosquito response to the combined visual and CO$_2$ cues. Plots correspond to Fig.~\ref{fig:stimulus}{\bf C} -- {\bf E}. 
    ({\bf E}) {\it Left}: Mosquito trajectories around a human subject wearing black hood and white outfit (shown in {\it Middle}). {\it Right}: The human subject can be approximated by a black sphere emitting CO$_2$. ({\bf F}) Heatmaps of mosquito densities at varying distances $d$ from the center of the head, flight speeds $v$, and flight directions $\hat{\mathbf{v}}\cdot\hat{\mathbf{d}}$, for the experiment in {\bf E}.
    ({\bf G}) The learned forces in {\bf C} are approximated by non-negative linear superposition, $f_{\parallel,\mathrm{lin}}$ and $f_{\perp,\mathrm{lin}}$, of the responses to individual cues. Heatmaps showing the difference between the inferred and linearly reconstructed forces indicate that the mosquito's response to combined stimuli is not a simple sum of uni-stimulus responses.
    ({\bf H}) Heatmaps of simulated trajectories predicted by the model learned in {\bf C}. Plots correspond to {\bf E}.
  }
    \label{fig:compare}
\end{figure*}

\begin{figure*}[!t]
    \centering
    \includegraphics[width = 0.65\textwidth]{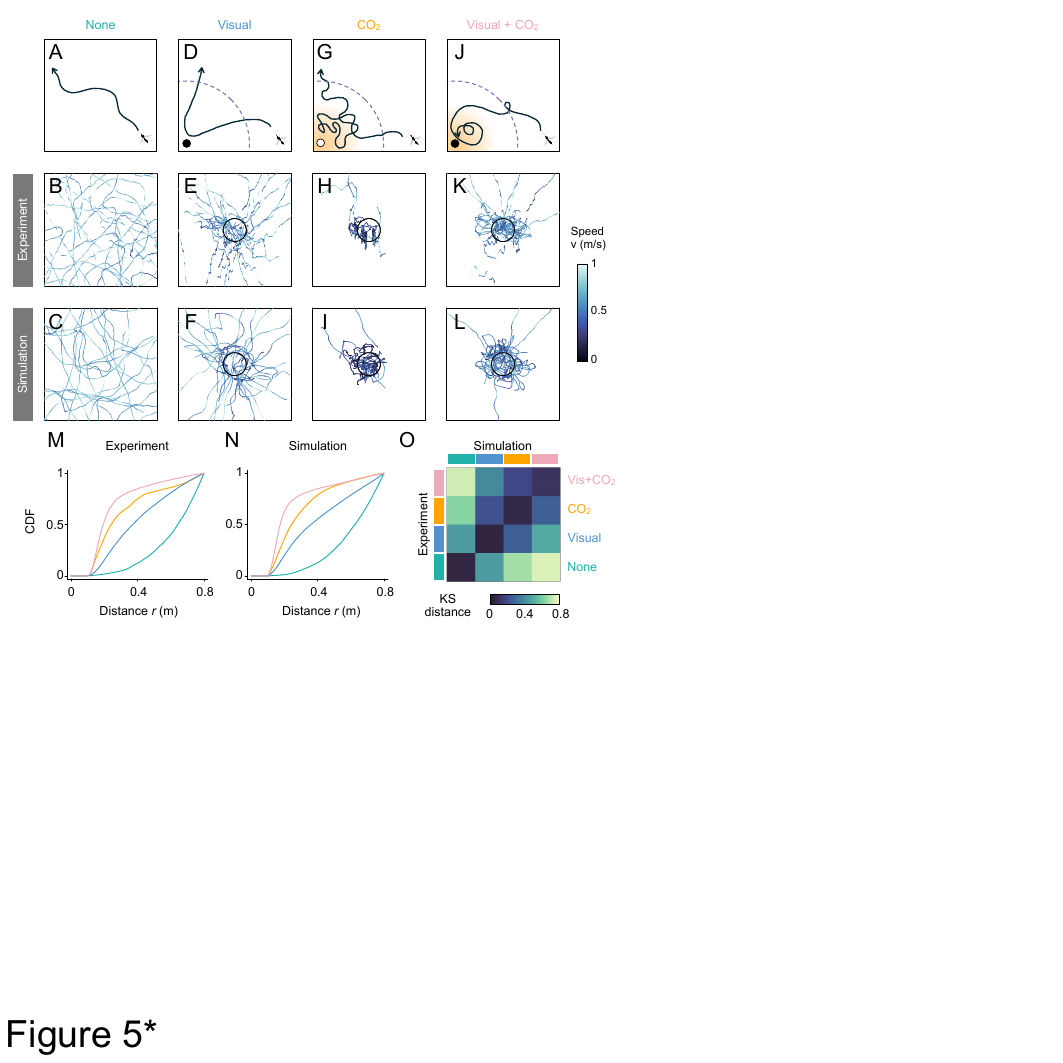}
    \captionsetup{labelsep = period, justification=justified,labelfont=bf,singlelinecheck = false}
    \caption[Model-predicted mosquito trajectories in response to different stimuli agree with experimental observations.]
    {
    {Model-predicted mosquito trajectories in response to different stimuli agree with experimental observations.} 
    ({\bf A}) Representative trajectory of a mosquito in an empty chamber. 
    %There is no cue present so the mosquito behavior is governed by noise.
    ({\bf B}) Typical experimental trajectories ($n$=30) in a 1~m by 1~m by 1~m cube without cues. These 3D trajectories are projected into a 2D plane.
    ({\bf C}) 2D projection of 3D simulated trajectories ($n=30$) of mosquitoes in an empty chamber.
    ({\bf D} -- {\bf L}) Schematic, experimental, and simulated trajectories of mosquitoes in response to ({\bf D} -- {\bf F}) visual cues, ({\bf G} -- {\bf I})
    CO$_2$ cues, and ({\bf J} -- {\bf L}) combined visual and CO$_2$ cues. %Plots correspond to {\bf A} -- {\bf C}. 
    For each cue, $n=30$ trajectories are shown. In panels {\bf D}, {\bf G}, and {\bf J}, dashed circles indicate the zone of attraction, and diffuse orange denotes CO$_2$. In panels {\bf H} and {\bf I}, the trajectories are spread out in the third dimension.
    ({\bf M}, {\bf N}) Cumulative distribution functions (CDFs) of ({\bf M}) the experimental trajectories and 
    ({\bf N}) the simulated trajectories in response to the specified sensory cues indicate that the learned models capture the key features of mosquito flight behavior. ({\bf O}) Kolmogorov–Smirnov (KS) distances demonstrate a quantitative agreement between the experimental and simulation CDFs.
  }
    \label{fig:trajectory}
\end{figure*}

How do mosquitoes respond to composite sensory stimuli? To explore this question, we track 3D mosquito trajectories around a black sphere releasing CO$_2$ plumes (Fig.~\ref{fig:compare}A), representing combined visual and CO$_2$ cues. The mosquito track densities are more concentrated around the spherical target compared to experiments with either visual or CO$_2$ cues alone (Fig.~\ref{fig:compare}B, left),  suggesting a stronger attraction to the combined stimulus. The mosquitoes were energetic: they maintained a flight speed of approximately 0.5 m/s near the target, higher than the speeds observed in the single-cue tests using CO$_2$ or visual cues. Moreover, our data show a high density of trajectories with $\hat{\mathbf{v}} \cdot \hat{\mathbf{d}}$ close to zero, indicating that mosquitoes were more likely to circulate around the target (Fig.~\ref{fig:compare}B, right). Individual trajectories further confirm the sustained orbiting behavior (Fig.~\ref{fig:trajectory}K and \SVid{6}). These results suggest that the combination of visual and CO$_2$ cues activates a distinct behavioral state that facilitates mosquito host-seeking. 

To better understand this behavioral state, we apply the same inference techniques as before to learn the $f_\parallel$ and $f_\perp$ forces in response to the combined cues. $f_\parallel$ shows signatures of responses to individual cues (Fig.~\ref{fig:compare}C). When the mosquitoes fly toward the target (negative $\hat{\mathbf{v}} \cdot \hat{\mathbf{d}}$), $f_\parallel$ mirrors the response to visual cues alone, representing a speed potential force with reduced speed near the target. When mosquitoes fly away from the target (positive $\hat{\mathbf{v}} \cdot \hat{\mathbf{d}}$), $f_\parallel$ resembles the response to CO$_2$ alone, showing a transition to extremely low speeds when the distance to the target is below 0.4 m. The $f_\perp$ force shows a similar pattern to that with visual cue alone, reorienting mosquitoes toward the target when far away while directing them away from the target when in close proximity, but with a much stronger magnitude (Fig.~\ref{fig:compare}C). 
Simulations of the inferred model quantitatively capture the density distributions of mosquito tracks observed in the experiment (Fig.~\ref{fig:compare}D), and reproduce the orbiting behavior around the target (Fig.~\ref{fig:trajectory}L and \SVid{7}).

To assess whether the response to combined cues could be understood by adding up the individual stimulus responses, we perform non-negative linear regression, approximating the learned force of the full response as a linear superposition $f_\mathrm{lin} = \alpha_1 f_\mathrm{visual} + \alpha_2 f_\mathrm{CO_2}$ (Fig.~\ref{fig:compare}G). We find that $f_{\parallel,\mathrm{lin}} = 0.63 f_{\parallel,\mathrm{visual}} + 0.39 f_{\parallel,\mathrm{CO_2}}$, consistent with the mixed features of individual responses in $f_\parallel$, and $f_{\perp,\mathrm{lin}} = 1.63 f_{\perp,\mathrm{visual}} + 0.00 f_{\perp,\mathrm{CO_2}}$, consistent with an amplified visual-driven reorientation pattern observed in $f_\perp$. However, the linear superposition of the force alone does not fully recapitulate the inferred forces in response to combined cues, with substantial deviations between the inferred and linearly reconstructed forces, particularly in regions near the target (Fig.~\ref{fig:compare}G and \SFig\ref{figS:linsup}). Moreover, simulations using the linearly combined forces fail to reproduce the experimentally observed trajectory density around targets with combined cues. These discrepancies indicate that mosquito responses to visual and CO$_2$ cues are not additive at the level of inferred behavioral forces, and may involve nonlinear integration and potential interactions between different sensory pathways \cite{alonso2022olfactory,van2015mosquitoes}. Indeed, CO$_2$ not only acts as an attractive bait for mosquitoes but also excites and activates them to be more responsive to other host-finding cues~\cite{mcmeniman2014multimodal,majeed2017detection}.

To further test the predictive power of the inferred model, we use our model to predict mosquito flight behavior around a human subject wearing all white except for a black hood, our best approximation of a ``spherical human" (Fig.~\ref{fig:compare}E). We approximate the human subject as a 12-inch-diameter (0.3 m) black sphere emitting CO$_2$, representing combined visual and CO$_2$ cues. The inferred behavioral force is rescaled to match the size of the human head (see Supplementary Materials). As shown in Figs.~\ref{fig:compare}F and H, model predictions quantitatively replicate the experimental mosquito densities around the human head, demonstrating the model's ability to accurately describe key features of mosquito behavior in a realistic setting.

\subsubsection*{Assessing mosquito bite risk}
To prevent the spread of disease, it is important to keep track of the potential bite risk, which we measure as the distance $d_{50}$ at which 50\% of mosquito trajectories are concentrated near an attractive target. Thus, a smaller $d_{50}$ value indicates a higher bite risk.  For an 8-inch spherical target, Fig.~\ref{fig:trajectory}M-N shows the cumulative distributions of mosquitoes as a function of distance from various targets, for both experimental data and simulations of the learned models. The $d_{50}$ values are roughly $0.65$ m for no cues, $0.4$ m for visual cues, $0.25$ m for CO$_2$ cues, and $0.2$ m for combined visual and CO$_2$ cues. The close agreement between the experimental and simulated data, as shown by the Kolmogorov–Smirno (KS) distance in Fig.~\ref{fig:trajectory}O, again highlights the accuracy of our model in predicting mosquito behavior under various conditions.

\subsection*{Discussion}

In this study, we developed protocols and an inference pipeline for quantitatively characterizing mosquito responses to host-linked cues, such as visual and CO$_2$ cues. To our knowledge, we have generated the largest mosquito 3D tracking dataset to date (\STable\ref{tabS:trajcompare} and \STable\ref{tabS:trajtotal}), which enabled the discovery of realistic dynamical models of mosquito behavior. We find that a combination of visual and CO$_2$ cues attracts more mosquitoes than individual cues alone. Our results provide valuable insights into mosquito flight patterns around human subjects. For instance, our experiments confirm that {\it Ae.~aegypti} mosquitoes show a preference for darker colors and primarily target the head of a human (Fig.~\ref{fig:demo}C and Fig.~\ref{fig:compare}E), which present both visual and CO$_2$ cues. To make these insights accessible to a broader audience, we have developed an interactive web application that incorporates all the learned mosquito models \cite{mosquitowebsite} (\SFig\ref{figS:website}). This application allows users to change the type of attractive cues and the number of mosquitoes, and to visualize mosquito flight dynamics.

Most previous experiments on mosquito host-finding have been conducted in wind tunnels where emitted CO$_2$ triggers mosquitoes to fly upwind towards the source~\cite{van2015mosquitoes,majeed2017detection,cooperband2006comparison,cooperband2006orientation}.  Our study of mosquitoes flying in a closed room is most similar to studies of mosquitoes flying downwind to hosts \cite{gillies1974evidence}, a situation which has received less attention.  In such conditions, odor and CO$_2$ cues stay relatively stationary, consistent with our observation that vision is an important part of host-finding.

While we primarily focused on understanding mosquito responses to a single target, we have also conducted preliminary experiments to explore how mosquitoes respond to multiple targets (\SFig\ref{figS:visualvarsizes} and \SFig\ref{figS:2sphere}). Developing a quantitative framework for how mosquitoes integrate cues from multiple targets and make decisions will be an important direction for future research. Our study of combined sensory cues provides a first step in this direction, laying the groundwork for understanding mosquito behavior in complex environments such as those involving human groups. Additionally, our current experimental setup focused on flight behavior, and the trajectories that we collected did not include landing and blood-feeding behaviors. Incorporating these aspects in future experimentation will be crucial for developing a comprehensive understanding of mosquito interactions with hosts.

We expect our framework to be widely applicable for quantifying the behavior of other mosquito species, such as the malaria mosquito {\it Anopheles gambiae}, as well as the behavioral changes caused by pathogen infections in mosquitoes and other vectors \cite{huang2023delftia,rivera2025rewiring}. These future studies could guide the development of effective prevention strategies for mosquito-borne diseases.

\subsection*{Acknowledgements}
These experiments are conducted under Georgia Institute of Technology's IRB protocol H23325 ensuring adherence to ethical standards and guidelines. 
The reagent was obtained through BEI Resources, NIAID, NIH: Aedes aegypti, Strain ROCK, MRA-734, contributed by David W. Severson

{\bf Funding:} This material was supported by the NSF Physics of Living Systems student network. This research received support through Schmidt Sciences, LLC (Polymath Award to J.D.) and the MIT MathWorks Professorship Fund (J.D.). A.E.C. was supported by the Department of Defense through the National Defense Science and Engineering Graduate Fellowship.

{\bf Author Contributions:}
R.T.C., J.D., and D.L.H.~designed research. C.Z.~and S.K.~performed the experiments. C.F.~and A.E.C.~performed modeling. C.F.~performed inference. A.E.C.~developed the web application. All authors analyzed the data. All authors wrote the paper.

{\bf Competing interests:} The authors declare no competing interests. 

{\bf Data and materials availability:} Data and codes are available online at \cite{mosquitobehaviordata} and \cite{mosquitocode}.

\clearpage
\bibliography{main-mosquito.bib}

% Main figures

% ====== Supp Info ======

\clearpage
\onecolumngrid
% \appendix
\begin{center}
\textbf{\large Supplementary Materials}
\end{center}

\renewcommand{\thesection}{\Roman{section}}
\renewcommand{\thesubsection}{\Alph{subsection}}
\renewcommand{\thetable}{\arabic{table}}
\renewcommand{\thefigure}{\arabic{figure}}
\renewcommand{\theequation}{S\arabic{equation}}

\renewcommand{\figurename}{Supplementary Fig.}
\renewcommand{\tablename}{Supplementary Table}

\setcounter{figure}{0}
\setcounter{equation}{0}
\setcounter{table}{0}
\setcounter{section}{0}

% Supplementary Notes
\section{Experimental materials and methods}

\subsection{Mosquito rearing}
In this study, we focus on the anthropophilic mosquito \textit{Aedes aegypti}, which feeds during the day and is a vector of several pathogens. 
%From CDC
The mosquito strain used in this study was obtained through BEI Resources, NIAID, NIH: {\it Aedes aegypti}, Strain ROCK, MRA-734, contributed by David W. Severson.
All mosquito rearing took place in an insect rearing room (Bahnson Environmental Chamber CCS-300, Clemmons, NC) at 27°C and 78\% humidity with 12:12 light and dark cycles that include 30-minute periods of sunrise/sunset. 

For each trial, a cohort of about five hundred mosquitoes were reared and then the female mosquitoes (7-21 days post emergence) were fed defibrinated rabbit blood (Hemostat Labs, DRB050, Dixon, CA) using a membrane feeder. Three days post-blood feeding, oviposition cups were placed inside each cage. The oviposition cup was a 150 ml cup lined with seed germination paper (Anchor Paper Company, SD7615L, St. Paul, MN) with purified water covering the bottom of the cup (depth 0.50 cm). Aedes species lay eggs on the inner wall of containers and just above the water line. By placing the seed germination paper around the inner wall of the cup, the mosquitoes lay eggs on the paper creating an egg sheet. Twenty-four hours later after placing it into the cage, the oviposition cup was removed. Excess water, eggs not adhered to the egg sheets, and dead adults were gently rinsed from the oviposition cup, and the cup with egg sheets was placed in a closed Tupperware container. It remained in the closed container for 72 hours and then the lid was cracked 1 cm to allow for water evaporation. Once the egg sheets were completely dry (48-72 hours later), egg sheets were removed from the cup and stored in plastic Ziploc bags. Aedes eggs can be stored in this manner for several months and still remain viable.

Two weeks prior to the scheduled experiment date, an egg sheet was removed from the Ziploc storage bag, placed in a petri dish and submerged in water and then placed in a vacuum chamber for 15 minutes at 15 psi. The egg sheet was then placed in a larval rearing pan (US Plastics, 52051, Lima, OH) with 250 ml purified water and 25 mg ground fish diet (Tetra, TetraMin Tropical Granules) to allow eggs to hatch and larvae to emerge. Larval rearing pans were placed in an insect rearing room (Bahnson Environmental Chamber CCS-300, Clemmons, NC) at the previously mentioned rearing conditions. Larvae were kept at a density of approximately 200 larvae per rearing pan and fed 4-6 pellets of high-quality koi food daily following standard Aedes rearing protocols. The larval pan water was cleaned daily. Pupae were put in a 150 ml plastic cup and placed in a Bug Dorm (MegaView Science, DP-1000, Taichung, Taiwan) with a hanging sugar vial (10\% sugar, 0.1\% methylparaben). Once all pupae had emerged into adults, the pupae cup was removed. The sugar vial was monitored daily to ensure it remained at a sufficient volume. Mosquitoes are stored in environmental chambers until the day of the experiment.
On the day of the experiment, mosquitoes are sorted manually and transferred from the storage cages into release capsules using mouth aspirators. 

\subsection{Capturing Trajectories}
To attain mosquito flight trajectories, we worked with the Centers for Disease Control and Prevention (CDC) in Atlanta to employ a state-of-the-art 3-D infrared camera system, the Photonic Fence Monitoring Device (PFMD). 
The PFMD system, developed by Photonic Sentry, is a specialized imaging device designed to detect and track flying insects with unprecedented precision and speed. The system uses a dual-lens camera, infrared LED's, and a retro-reflective backdrop for precise position tracking of mosquitoes at time steps of 0.001 seconds \cite{keller2020optical}. 

After obtaining trajectories from the PFMD, we filter to remove early time trajectories to mitigate effects from the mosquito release, short time trajectories, and trajectories near the walls of the chamber to eliminate effects from mosquitoes landing on walls. 
To eliminate effects from mosquito release, we remove all trajectories within the first 5 minutes of the experiment.
Furthermore, we remove all trajectories that are shorter than 1.0 second, as suggested by the PFMD manufacturers, to reduce noisy trajectories. 
Finally, to remove trajectories from mosquitoes landing on the walls, we identify the walls by finding the 1st and 99th percentile of all mosquito positions in each dimension, and we remove mosquitoes within 10 cm of the side walls, within 10 cm of the top wall and floor, and with 50 cm of the front and back walls.

\subsection{Experimental Setup}
For the experiment, a chamber was set up with mesh netting extending 502~cm from a 162~cm by 152~cm clear acrylic panel to a 312~cm by 314~cm retro-reflective sheet, creating a large trapezoidal prism shown in main Fig.~1. The PFMD camera is set up across the room adjusting the focal length and the infrared LED brightness to fully illuminate the contained trapezoidal prism. A plastic tarp covers the chamber to minimize external flows in the chamber. The environmental chamber is set at 28 degrees Celsius and 45 percent humidity. Tests are conducted with 50 or 100 mosquitoes and last 10 to 20 minutes.
% The experiments are performed by placing various sensory cues inside the chamber. 

% Free flight behaviors
\noindent {\bf Free flight}:
To understand mosquito behavior in the environmental chamber, mosquitoes are aspirated from their holding container (container with food) into a release canister and visually confirmed to be female. They are then released from the canister by knocking over a cylindrical canister (3 in tall 1 in diameter) and removing its lid. The initial shock ensures that a large majority of the mosquitoes fly out of the capsule. The release into an empty chamber and their subsequent behavior is recorded with the PFMD camera. To characterize the empty chamber, experiments were conducted with 100 mosquitoes for 20 min. These datasets are used to validate our inference framework for learning dynamical models for mosquito flight behaviors (see main Fig.~2 and Supplementary Fig.~\ref{figS:levitation}).

%Black Ball 6in suspended ~5 ft off the ground
\noindent {\bf Human mimic experiments}: 
To introduce more cues, we performed experiments with styrofoam spheres as visual cues, with and without emitting carbon dioxide.

We used 4, 8, 12, and 16 inch-diameter styrofoam spheres to determine how behavior depends on target size. The balls were either painted black or kept bare (white).  For all experiments, the ball was elevated 5~ft off the ground, which is the midpoint between the ceiling and floor. The target is placed 300 cm away from the mosquito release point. (800 cm away from the camera)

To further understand how the visual cue affects the mosquito behavior a preference test is conducted. Sphere preference testing is conducted by creating a T joint on top of the 5~ft stand allowing for the placement of two spheres on either side of the stand. Tests were conducted with two 4~in spheres, a 4 and 8~in sphere, a 4 and 12~in sphere, and a 4 and 16~in sphere.

To study the effect of carbon dioxide, tests were conducted with spheres atop a carbon dioxide release. The carbon dioxide is released from a CO$_2$ tank and controlled with a calibrated flow meter. A volumetric flow rate of 0.24 L/min is released from the 0.25~in diameter tube. The flow rate matches the rate of a human breathing. The CO$_2$ experiment is conducted with the 8~in sphere as it gave the best results from the single sphere experiments.

\noindent {\bf Human experiment}: 
Humans are an optimal bait source as many mosquito species have evolved specifically to target humans\cite{mcbride2014evolution}. Using the PFMD camera and a human target, a clear silhouette is created with the mosquito trajectories around the target in the dark clothing experiment. The target is posed as shown in Fig 1 B and C where each limb can be clearly seen in a silhouette. 

We perform 3 experiments with the same human subject, but each wearing different outfits.  In the first experiment the target is wearing a dark gray fleece sweatshirt, blue jeans, cotton socks, and dark colored lab gloves, emulating a human target wearing dark colored clothing. This outfit is shown in Fig.~1B.
In the second experiment, the human target wears clothes such that one side is white and the other is black allowing for visual asymmetry shown in Fig.~1C. The top is a fleece sweatshirt where a cotton t-shirt is sewn into the hood to match the white side of the sweatshirt. The bottom is a pair of jeans where half of it is white and the other half is black. 
Finally, in the white torso and black head experiment, the hood is from a black fleece sweatshirt where the torso of the jacket is covered by a white polyester jacket. For the bottom the human target is wearing white fleece sweatpants and white shoes. This outfit is shown in Fig.~4E.
In each of these experiments precautions were taken to protect the target's face from mosquito bites by sewing/taping a white polyester mosquito mesh to the front opening of the sweatshirt hood. 
To run the experiments the human target gets into position at the 800~cm mark from the camera and uses a long white PVC to knock down the release capsule in doing so releasing the mosquitoes. The human target will stand still in their pose to allow for clear tracking of mosquitoes and the formation of a silhouette. 

\section{Mosquito swarming analysis}
\begin{figure*}[!t]
    \centering
    \includegraphics[width = .8\textwidth]{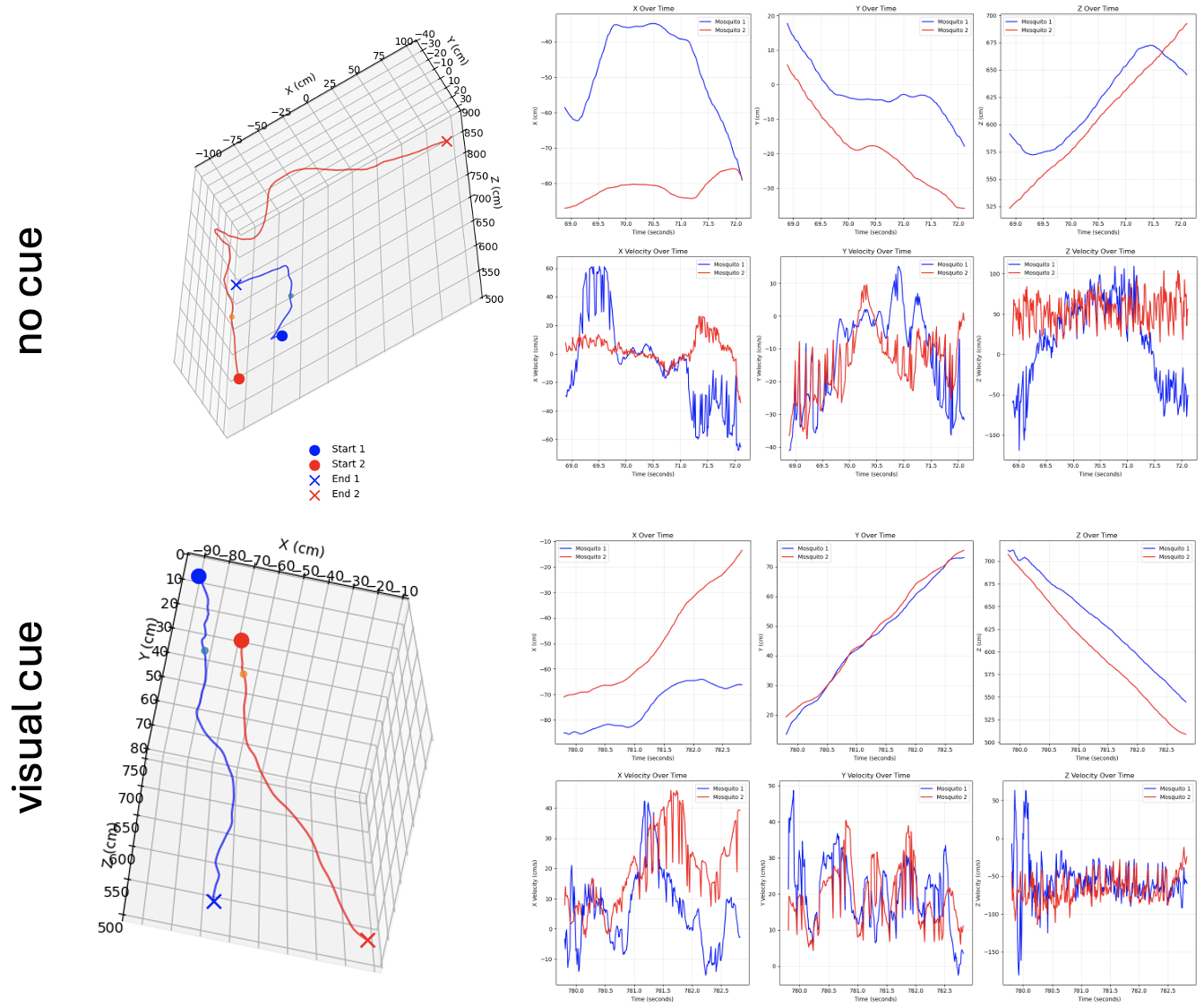}
    \captionsetup{labelsep = period, justification=justified,labelfont=bf,singlelinecheck = false}
    \caption[]
    {
    Example pair mosquito trajectories that follow similar paths for a brief period of time.
  }
    \label{figS:pair_traj}
\end{figure*}

While we did not find any evidence of mosquito swarming behavior in the data, we did find a small number of trajectories in each dataset in which pairs of mosquitoes follow very similar paths (\SFig\ref{figS:pair_traj}).
These trajectories only occurred for mosquito pairs and numbered in the 1-10 range out of 10,000s.
Currently, we do not know the cause of these trajectories but we believe they are due to mosquitoes avoiding each other or mosquitoes having similar behavioral reactions to cues.

\section{Bayesian inference of dynamical models \label{sec:models}}
To describe mosquito flight behavior, we use the following equations to model the temporal changes in mosquito positions $\mathbf{r}$ and flight velocities ${\mathbf{v}}$:
\begin{equation}
\dot{\mathbf{r}} = \mathbf{v}, ~~\dot{\mathbf{v}} = \mathbf{f}(\mathbf{v}, \mathbf{r}) + \boldsymbol{\xi}, \label{eqn:newtonian}
\end{equation}
where the dot symbol ($\dot{~}$) denotes time derivative, and $\mathbf{f}(\mathbf{v},\mathbf{r})$ is the total forces that mosquitoes experience during their flight. The Gaussian white noise $\boldsymbol{\xi}$ satisfies $\boldsymbol{\xi}(t)\boldsymbol{\xi}(t^\prime) = \Delta \mathbf{I} \delta(t-t^\prime)$, where $\Delta$ is the magnitude of the noise, $\mathbf{I}$ is an identity matrix of size 3, and $\delta$ denotes the Kronecker delta function. Our goal is to infer the behavioral forces of mosquitoes directly from 3D time-series measurements of mosquito trajectories $\mathbf{r}(t)$.

\subsection{Representation of mosquito behavioral forces}
To model the total forces of free-flying mosquitoes in the absence of sensory cues, we decompose $\mathbf{f} = \alpha(v)\mathbf{v} + f_z(v, \hat{\mathbf{v}}\cdot\hat{\mathbf{z}}) \hat{\mathbf{z}} - g\hat{\mathbf{z}}$ into three different components: a thrust force $\alpha(v)\mathbf{v}$ in the direction of velocity $\mathbf{v}$, a levitation force $f_z(v, \hat{\mathbf{v}}\cdot\hat{\mathbf{z}}) \hat{\mathbf{z}}$, and a gravitational force $-g\hat{\mathbf{z}}$ with a constant $g\approx10~\mathrm{m/s^2}$. Here, we use the hat symbol ($\hat{~}$) to denote unit vectors, and we use non-bold letters to denote the magnitude of a vector. We note that the levitation force is assumed to be independent of the absolute height $z$. The dot product $\hat{\mathbf{v}}\cdot\hat{\mathbf{z}}$ describes the direction of mosquito flight, with positive values indicating upward flight and negative values indicating downward flight. We approximate $\alpha(v)$ by a linear combination of basis functions
\begin{equation}
\alpha(v) = \sum_{m} w_m \theta_m(v;v_0), \label{eqn:bf_thrust}
\end{equation}
where $\theta_m = L_m(v/v_0) \exp(\frac{-v}{2v_0})$ represents Laguerre polynomials $L_m$ with exponential weighting factors, and the weights $w_m$ encode the information about $\alpha(v)$ that we aim to learn from data. Similarly, we expand the magnitude of the levitation force $f_z(v, \hat{\mathbf{v}}\cdot\hat{\mathbf{z}}) \hat{\mathbf{z}}$ as 
\begin{equation}
f_z(v, \hat{\mathbf{v}}\cdot\hat{\mathbf{z}}) = \sum_\mu w_\mu \vartheta_\mu(v, \hat{\mathbf{v}}\cdot\hat{\mathbf{z}}) = \sum_{\mu_0}\sum_{\mu_1} w_{(\mu_0, \mu_1)} {\theta}_{\mu_0}(v; v_0) {\theta}_{\mu_1}^\ast(\hat{\mathbf{v}}\cdot\hat{\mathbf{z}}),  \label{eqn:bf_levitation}
\end{equation}
where each basis function $\vartheta_\mu$ is decomposed into the product of univariate functions $\theta$ and $\theta^*$. Here, $\theta$ represents the same weighted Laguerre polynomials as described above, and $\theta^\ast$ denotes the Legendre polynomials. As shown in \SFig\ref{figS:levitation}, the learned levitation force has a constant magnitude $f_z = g$ that balances the gravitational force. Therefore, we ignore these two components in our analysis below. 

To model the behavioral forces that mosquitoes use to maneuver their flight in response to environmental stimuli, we consider point-source stimuli and decompose the force $\mathbf{f} = f_\parallel \hat{\mathbf{v}} + f_\perp (\mathbf{I} - \hat{\mathbf{v}}\hat{\mathbf{v}})\cdot\hat{\mathbf{d}} $ into two orthogonal components (see main Fig.~3): one parallel to the direction of flight $\hat{\mathbf{v}}$, and one perpendicular to it. Similar to the procedures above, we represent the force magnitudes $f_\parallel$ and $f_\perp$ using basis-function expansions:
\begin{subequations}
\label{eqn:bf_response}
    \begin{equation}
        f_\parallel(v, d, \hat{\mathbf{v}}\cdot\hat{\mathbf{d}}) = \sum_{\mu_0, \mu_1, \mu_2} w^\parallel_{(\mu_0, \mu_1, \mu_2)} {\theta}_{\mu_0}(v; v_0)  {\theta}_{\mu_1}(d; d_0) {\theta}_{\mu_2}^\ast(\hat{\mathbf{v}}\cdot\hat{\mathbf{d}}), \label{eqn:bf_response_para}
    \end{equation}
    \begin{equation}
        f_\perp(v, d, \hat{\mathbf{v}}\cdot\hat{\mathbf{d}}) = \sum_{\mu_0, \mu_1, \mu_2} w^\perp_{(\mu_0, \mu_1, \mu_2)} {\theta}_{\mu_0}(v; v_0)  {\theta}_{\mu_1}(d; d_0) {\theta}_{\mu_2}^\ast(\hat{\mathbf{v}}\cdot\hat{\mathbf{d}}), \label{eqn:bf_response_perp}
    \end{equation}
\end{subequations}
where $\mathbf{d}$ denotes the displacement vector from the point stimulus to the mosquitoes.

\subsection{Sparse Bayesian inference}
To infer the behavioral forces directly from experiments, we first compute $\mathbf{v}$ and $\dot{\mathbf{v}}$ from the measurements of $\mathbf{r}(t)$ by using finite-difference methods. This allows us to construct the basis functions described in Eqs.~(\ref{eqn:bf_thrust})-(\ref{eqn:bf_response}). Introducing these expressions into Eq.~(\ref{eqn:newtonian}) and stacking all the trajectories into a matrix form, we obtain $\dot{\mathbf{v}} = \boldsymbol{\Theta}\mathbf{w} + \boldsymbol{\xi}$, which reduces the inference task into a linear regression problem. Here, each row corresponds to a spatial dimension of a mosquito trajectory at a specific time point, each column of $\boldsymbol{\Theta}$ corresponds to a mode of the behavioral forces in Eqs.~(\ref{eqn:bf_thrust})-(\ref{eqn:bf_response}), and all the coefficients in the basis-function expansion are grouped in a coefficient vector $\mathbf{w}$.

To perform Bayesian inference of $\mathbf{w}$, we minimize the negative log-posterior
\begin{equation}
-\ln P(\mathbf{w} | \{\mathbf{r}(t)\} ) \sim -\ln P(\{\mathbf{r}(t)\} | \mathbf{w}) - \ln P(\mathbf{w}) \label{eqn:log_posterior}
\end{equation}
with respect to $\mathbf{w}$ given the measurements $\{\mathbf{r}(t)\}$. To prevent overfitting, we follow previous work \cite{tipping2001sparse, wipf2004sparse} and impose a sparsity-promoting Gaussian prior over the coefficients
\begin{equation}
P(\mathbf{w})
= \prod_{m} \mathcal{N}(w_m | 0, \gamma_m)
= \prod_{m} (2\pi \gamma_m)^{-1/2} \exp\Big( - \frac{w_m^2}{2\gamma_m}\Big), 
\end{equation}
where $\gamma_m$ are hyperparameters representing the variances of the Gaussian distributions. Therefore, the negative log-prior becomes
\begin{equation}
-\ln P(\mathbf{w}) = \sum_{m} \frac{w_m^2}{2\gamma_m} + \sum_{m} \frac{1}{2}\ln(2\pi\gamma_m), \label{eqn:log_prior}
\end{equation}
which is similar to an L2 regularization on the coefficients $\mathbf{w}$ \cite{krogh1991simple, mackay1992practical}.
The negative log-likelihood function is thus given by 
\begin{equation}
-\ln P(\{\mathbf{r}(t)\}|\mathbf{w}) =  \frac{N}{2}\ln(2\pi) + \frac{1}{2}\ln |\boldsymbol{\Psi}| +  \frac{1}{2}(\mathbf{v}-\boldsymbol{\Theta}\mathbf{w})^T\boldsymbol{\Psi}^{-1}(\mathbf{v}-\boldsymbol{\Theta}\mathbf{w}), \label{eqn:log_likelihood}
\end{equation}
where $\boldsymbol{\Psi} = \Delta \mathbf{I}_N$ is a diagonal matrix assuming Gaussian white noise, and $N$ is the number of rows $\boldsymbol{\Theta}$ contains. Introducing Eqs.~(\ref{eqn:log_prior}, \ref{eqn:log_likelihood}) into Eq.~(\ref{eqn:log_posterior}), we obtain the posterior that follows a Gaussian distribution $P(\mathbf{w} | \{\mathbf{r}(t)\}) = \mathcal{N}(\mathbf{w};\boldsymbol{\mu}, \boldsymbol{\Sigma})$, where the covariance matrix $\boldsymbol{\Sigma}$ and the mean $\boldsymbol{\mu}$ are given by
\begin{gather}
\boldsymbol{\Sigma} = (\boldsymbol{\Theta}^T \boldsymbol{\Psi}^{-1}\boldsymbol{\Theta}^T + \boldsymbol{\Gamma}^{-1})^{-1} \label{eqn:Sigma}\\
\boldsymbol{\mu} = \boldsymbol{\Sigma}\boldsymbol{\Theta}^T\boldsymbol{\Psi}^{-1} \mathbf{v}, \label{eqn:mu}
\end{gather}
and $\boldsymbol{\Gamma}$ is a diagonal matrix with $\boldsymbol{\Gamma}_{mm}= \gamma_m$.

To determine the values of hyperparameters in Eqs.~(\ref{eqn:Sigma}, \ref{eqn:mu}), we employ a pragmatic procedure based on previous work \cite{mackay1992bayesian}, and choose $\gamma_m$ and $\Delta$ to maximize the marginal likelihood $P(\{\mathbf{r}(t)\} | \Delta, \{\gamma_m\}) = \int P(\{\mathbf{r}(t)\} | \mathbf{w}; \Delta) P(\mathbf{w} | \{\gamma_m\})  d\mathbf{w}$. We use the Expectation Maximization (EM) method to iteratively update the values of $\gamma_m$ and $\Delta$. Specifically, given $\gamma_m^{(n)}$ and $\Delta^{(n)}$ from the previous iteration, we compute the current estimate of $\boldsymbol{\mu}^{(n)}$ and $\boldsymbol{\Sigma}^{(n)}$ using Eqs.~(\ref{eqn:Sigma}, \ref{eqn:mu}). The EM approach gives the re-estimates
\begin{gather}
\gamma_m^{(n+1)} = \mathrm{E}_{\mathbf{w}\sim\mathcal{N}(\boldsymbol{\mu}^{(n)}, \boldsymbol{\Sigma}^{(n)})}[\mathbf{w}_m^2]=(\boldsymbol{\mu}_m^{(n)})^2 + \boldsymbol{\Sigma}_{mm}^{(n)}),~\mathrm{and} \label{eqn:update_mu}\\
\Delta^{(n+1)} = \mathrm{E}_{\mathbf{w}\sim\mathcal{N}(\boldsymbol{\mu}^{(n)}, \boldsymbol{\Sigma}^{(n)})}\Big[\frac{|\mathbf{v}-\boldsymbol{\Theta}\mathbf{w}|^2}{N}\Big] = \frac{1}{N}\Big[|\mathbf{v}-\boldsymbol{\Theta}\boldsymbol{\mu}^{(n)}|^2 + \Delta^{(n)}\sum_{m}(1-\boldsymbol{\Sigma}_{mm}^{(n)}/\gamma_m^{(n)}) \Big]. \label{eqn:update_Detla}
\end{gather}
We note that when the degrees of freedom of the data samples $N$ is much larger than the number of modes $M$, Eq.~(\ref{eqn:update_Detla}) can be approximated by $\Delta^{(n+1)}\approx |\mathbf{v}-\boldsymbol{\Theta}\boldsymbol{\mu}^{(n)}|^2 / N$. Finally, we note that the scale parameters $v_0$ and $d_0$ in Eqs.~(\ref{eqn:bf_thrust})-(\ref{eqn:bf_response}) are chosen to minimize the residual $|\mathbf{v}-\boldsymbol{\Theta}\boldsymbol{\mu}^{(n)}|^2 / N$ through an additional layer of optimization outside of EM. Once the response to a specific target is learned, we can rescale the $d_0$ parameters to approximate the response to a larger or smaller target.

\subsection{Model selection and verification}
\begin{figure*}[!t]
    \centering
    \includegraphics[width = 0.9\textwidth]{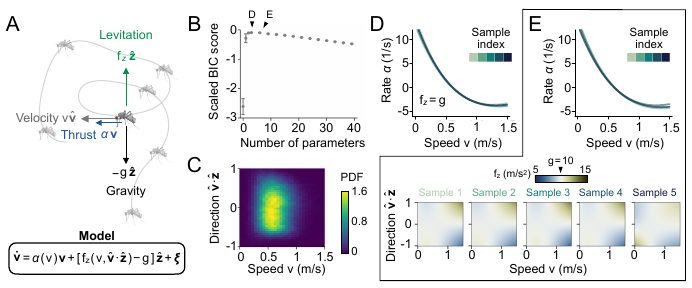}
    \captionsetup{labelsep = period, justification=justified,labelfont=bf,singlelinecheck = false}
    \caption[Dynamical inference of mosquito free-flight behaviors indicates the balance between levitation and gravity.]
    {
    {Dynamical inference of mosquito free-flight behaviors indicates the balance between levitation and gravity.} 
    ({\bf A}) Illustration of a mosquito flying with a velocity ${\bf v} = v\hat{\bf v}$. The (normalized) forces exerted on a mosquito include the thrust force $\alpha \mathbf{v}$, the levitational force $f_z \hat{\bf z}$, and the gravitational force $-g\hat{\bf z}$, where $g \approx 10~\mathrm{m/s^2}$ is the gravitational acceleration constant.
    ({\bf B}) Bayesian information criteria (BIC) scores for models with varying sparsity (number of model parameters). Arrowheads indicate two models shown in {\bf D} and {\bf E}. Error bars denote standard deviation based on 5 random subsamples, each containing 30 \% of the total trajectories.
    ({\bf C}) Probability density of recorded free flying mosquito trajectories at varying flight speed $v$ and direction $\hat{\bf v} \cdot \hat{\bf z}$. $\hat{\bf v} \cdot \hat{\bf z} = 1$ and $-1$ represent flying upward and downward, respectively.
    ({\bf D}) The inferred rate $\alpha$ of the model with the highest BIC score. This model has a constant levitational force $f_z = g$. Color scale indicates inference results for different subsamples.
    ({\bf E}) The model with the second highest BIC score shows inferred rates $\alpha$ highly similar to those in {\bf D}, and levitational forces $f_z$ approximately equal to $g$ in the high data density region in {\bf C}. 
  }
    \label{figS:levitation}
\end{figure*}

To generate models with varying complexities, we apply thresholding to the coefficients $\mathbf{w}$ and shrink the columns in $\boldsymbol{\Theta}$ sequentially. To identify the model that best balances the goodness of fit and model complexity, we use the negative Bayesian information criterion (BIC) \cite{ghosh2006introduction}, and select the model with the highest BIC score (see for example \SFig\ref{figS:levitation}B).

To verify the inferred model for mosquito flight behaviors, we simulate Eq.~(\ref{eqn:newtonian}) with the learned behavioral forces and noises, and the experimental initial conditions. We compare the relevant statistics between the simulated and experimental trajectories. The stochastic differential equations are solved using Julia's DifferentialEquations.jl package \cite{DifferentialEquations.jl-2017} with the SOSRA solver.
% In the main text, we primarily use spectral decompositions to represent the the velocity potential and mosquito interaction potentials.

\subsection{Applications to synthetic data \label{sec:synthdata}}

\begin{figure*}[!t]
    \centering
    \includegraphics[width = 0.8\textwidth]{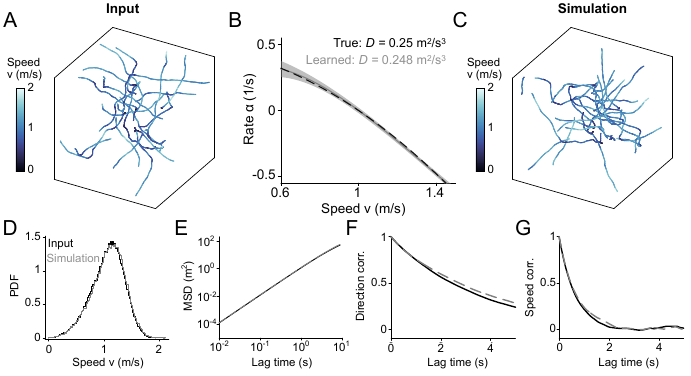}
    \captionsetup{labelsep = period, justification=justified,labelfont=bf,singlelinecheck = false}
    \caption[Bayesian inference framework accurately learns the Langevin equation with a speed potential force from synthetic data.]
    {
    {Bayesian inference framework accurately learns the Langevin equation with a speed potential force from synthetic data.} 
    ({\bf A}) Typical trajectories (20\%) of the input synthetic data for the Bayesian inference framework. 
    ({\bf B}) The inferred rate $\alpha$ (gray) accurately captures the ground truth (black). Shaded band indicates standard deviations computed from the posterior. See Sec.~\ref{sec:synthdata} for details.
    ({\bf C}) Typical simulated trajectories (20\%) of the learned model.
    ({\bf D} -- {\bf G}) The simulation of the learned model and the input data show quantitative agreement in:({\bf D}) probability distribution of flight speed, ({\bf E}) Mean squared displacement, ({\bf F}) directional correlation, and ({\bf G}) speed correlation. 
   }
    \label{figS:demo_velpot}
\end{figure*}

Before applying our inference framework on experimental data, we test our inference framework on synthetic data of simulated mosquito trajectories. To mimic mosquitoes' free flight, we first simulate the Langevin equation Eq.~(\ref{eqn:newtonian}) with a speed potential force $\mathbf{f} = \alpha(v; v_0) \mathbf{v}$. Here, we choose a quadratic function for the rate $\alpha(v, v_0) = \beta(v_0^2 - v^2)$ where $\beta=0.5~\mathrm{s/m^2}$ is the prefactor and $v_0 = 1~\mathrm{m/s}$ is the target flight speed. Indeed, simulations of this model yield trajectories that resemble the experimental data (see main Fig.~2), with a characteristic persistence length and moderate stochasticity in flight speed (\SFig\ref{figS:demo_velpot}A). We take these simulated trajectories as input data for the inference framework. The learned rate $\alpha(v; v_0)$ has a sparse representation with only a few nonzero coefficients and agrees quantitatively with the ground-truth function (\SFig\ref{figS:demo_velpot}B). Specifically, the Bayesian framework allow us to estimate the inference error of $\alpha(v; v_0)$ as  $\sigma_\alpha^2(v) = \boldsymbol{\theta}(v)^T \boldsymbol{\Sigma} \boldsymbol{\theta}(v)$ where $\boldsymbol{\Sigma}$ is the covariance matrix given by Eq.~(\ref{eqn:Sigma}) and $\boldsymbol{\theta} = [\theta_1(v), \theta_2(v),\dots, \theta_m(v), \dots ]^T$ is a vector of basis functions. To further validate the inferred model, we re-simulate Eq.~ (\ref{eqn:newtonian}) with the learned $\mathbf{f}$ to generate new trajectories of mosquito flight (\SFig\ref{figS:demo_velpot}C), and compare their statistics with those of input data. Specifically, we examined the speed distribution, mean squared displacement, direction correlation, and speed correlation of the re-simulated trajectories, finding a close match with the input data (\SFig\ref{figS:demo_velpot}D--G).

\begin{figure*}[!t]
    \centering
    \includegraphics[width = 0.8\textwidth]{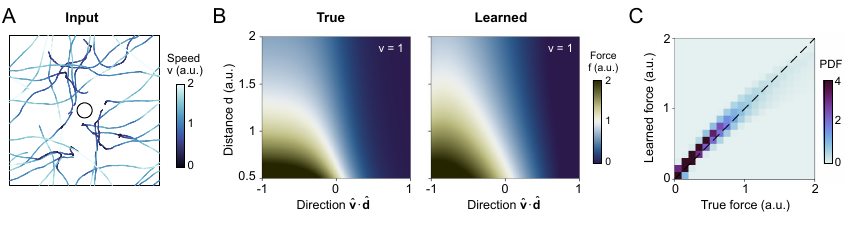}
    \captionsetup{labelsep = period, justification=justified,labelfont=bf,singlelinecheck = false}
    \caption[Bayesian inference framework accurately learns the response to external cues from synthetic data.]
    {
    {Bayesian inference framework accurately learns the response to external cues from synthetic data.} 
    ({\bf A}) Typical trajectories (10\%) of the input synthetic data for the Bayesian inference framework. 
    ({\bf B}, {\bf C}) The ({\bf} B)true and ({\bf C}) inferred behavioral force $f$ for the specified speed $v$. See Sec.~\ref{sec:synthdata} for details.
    ({\bf C}) Density heatmap showing the true force versus the learned force for varying values of flight speed $v$, flight direction $\hat{\mathbf{v}}\cdot\hat{\mathbf{d}}$, and distance toward the target $d$. The dashed line indicates the diagonal $y=x$.
    % are compared between experiments (black) and simulations of the learned model (gray).
   }
    \label{figS:demo_response}
\end{figure*}

\begin{figure*}[!t]
    \centering
    \includegraphics[width = 0.6\textwidth]{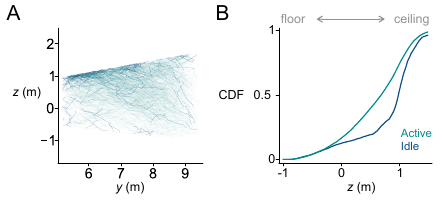}
    \captionsetup{labelsep = period, justification=justified,labelfont=bf,singlelinecheck = false}
    \caption[Mosquitoes are more likely to be idle than active when near the ceiling.]
    {
    {Mosquitoes are more likely to be idle than active when near the ceiling.} 
    ({\bf A}) Side-view trajectories of free-flying mosquitoes. The Idle state is colored blue and the active state is colored light green.
    ({\bf B}) Cumulative probability distribution of the two states as a function of vertical coordinate $z$. Larger $z$ values indicate being closer to the ceiling.
  }
    \label{figS:zcdf_active_idle}
\end{figure*}

\begin{figure*}[!t]
    \centering
    \includegraphics[width = .8\textwidth]{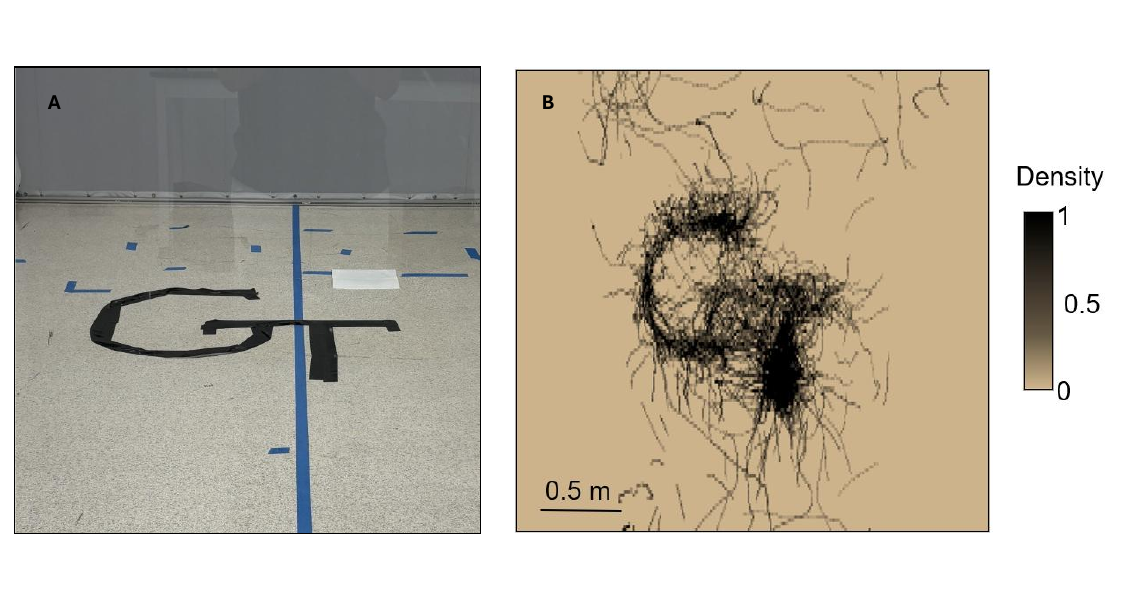}
    \captionsetup{labelsep = period, justification=justified,labelfont=bf,singlelinecheck = false}
    \caption[]
    {
    {Mosquito trajectories form the GT Logo} 
    ({\bf A}) The dark tape on the ground in the form of the GT Logo
    ({\bf B}) Mosquito trajectory concentration forming the Logo
  }
    \label{figS:logo}
\end{figure*}

\begin{figure*}[!t]
    \centering
    \includegraphics[width = .8\textwidth]{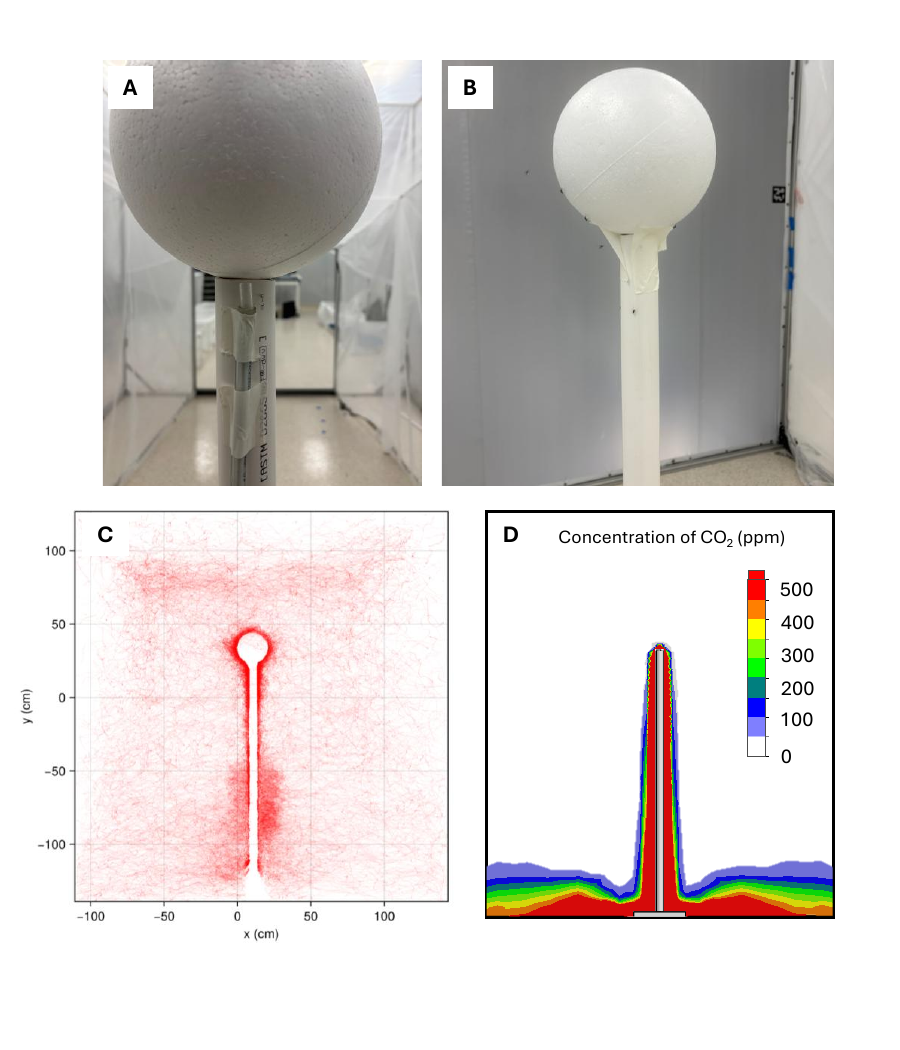}
    \captionsetup{labelsep = period, justification=justified,labelfont=bf,singlelinecheck = false}
    \caption[]
    {
    {CO$_2$ experimental setup.} 
    ({\bf A}) The setup of experiment where the CO$_2$ is released into the chamber.
    ({\bf B}) Mosquitoes landing on the white sphere and stand.
    ({\bf C}) Trajectories of mosquitoes around the white sphere and stand.
    ({\bf D}) Concentration of CO$_2$ above ambient air when released from the setup shown in {\bf A}.
  }
    \label{figS:co2}
\end{figure*}

\begin{figure*}[!t]
    \centering
    \includegraphics[width = 0.95\textwidth]{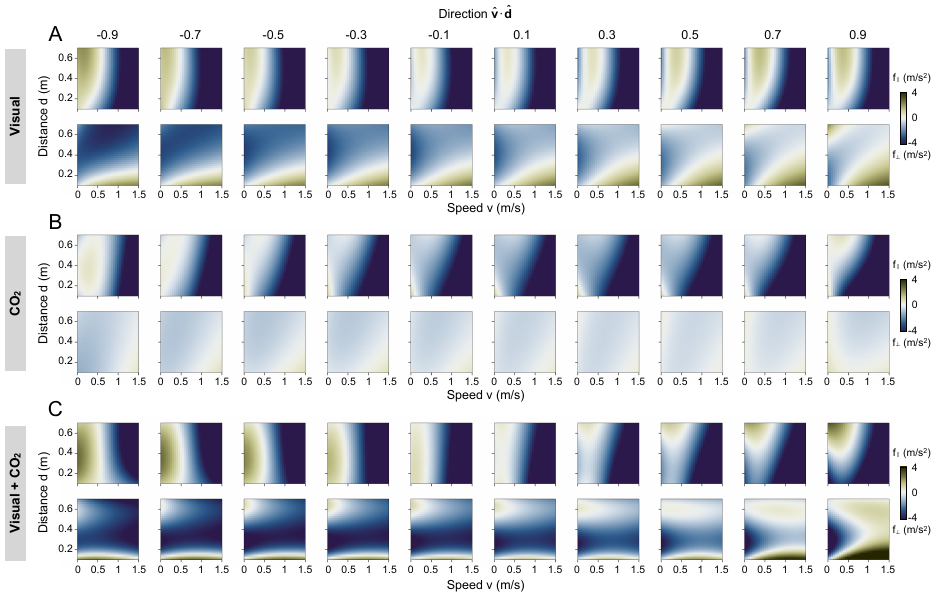}
    \captionsetup{labelsep = period, justification=justified,labelfont=bf,singlelinecheck = false}
    \caption[Learned behavioral forces demonstrate differential mosquito responses to various sensory cues.]
    {
    {Learned behavioral forces demonstrate differential mosquito responses to various sensory cues.} 
    ({\bf A}-{\bf C}) Learned behavioral forces acting parallel ({\it top}, $f_\parallel$) and perpendicular to ({\it bottom}, $f_\perp$) the mosquito flight direction, in response to ({\bf A}) visual cues, ({\bf B}) CO$_2$ cues, and ({\bf C}) combined visual and CO$_2$ cues.
  }
    \label{figS:learnedforces}
\end{figure*}

\begin{figure*}[!t]
    \centering
    \includegraphics[width = 0.6\textwidth]{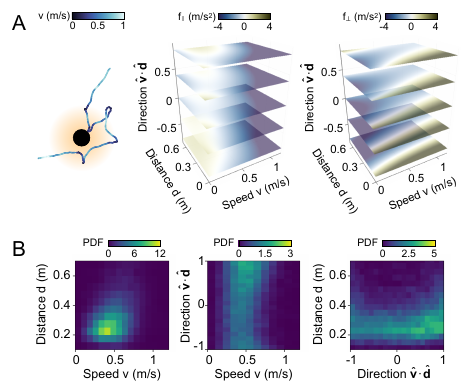}
    \captionsetup{labelsep = period, justification=justified,labelfont=bf,singlelinecheck = false}
    \caption[Mosquito response to combined visual and CO$_2$ cues is not a linear superposition of its response to individual cues.]
    {
    {Mosquito response to combined visual and CO$_2$ cues is not a linear superposition of its response to individual cues.} 
    ({\bf A}) {\it Left}: A typical 2D simulation trajectory of the model using a linear superposition of uni-stimulus response (see main text for details). 
    Heatmaps of the linearly reconstructed forces ({\it Middle}) $f_\parallel$ and ({\it Right}) $f_\perp$. ({\bf B}) 2D density maps of simulated mosquito trajectories using the behavioral forces in {\bf A} fail to capture the experimental data in main Fig.~4F, indicating that the mosquito's response to combined visual and CO$_2$ cues is not a linear superposition of its response to individual cues.
  }
    \label{figS:linsup}
\end{figure*}

Next, we test our inference framework on a simulation model of sensory response. We consider a repulsive force $\mathbf{f} = \gamma \frac{v \hat{\mathbf{d}} }{1+\exp(\lambda(\hat{\mathbf{d}} \cdot \hat{\mathbf{v}}))}$ in response to a point source stimulus. Here, we choose $\gamma =1$ and $\lambda = 5$. The simulated trajectories are shown in \ref{figS:demo_response}A. We use the basis-function expansion in Eq.~(\ref{eqn:bf_response}) and apply sparse Bayesian inference to learn the force magnitude. The close agreement between the learned force  and the true force (\SFig\ref{figS:demo_response}B--D)demonstrates that our inference framework is effective in learning the behavioral force from time-series trajectories. 

\section{Interactive Website}
\begin{figure*}[!t]
    \centering
    \includegraphics[width = .8\textwidth]{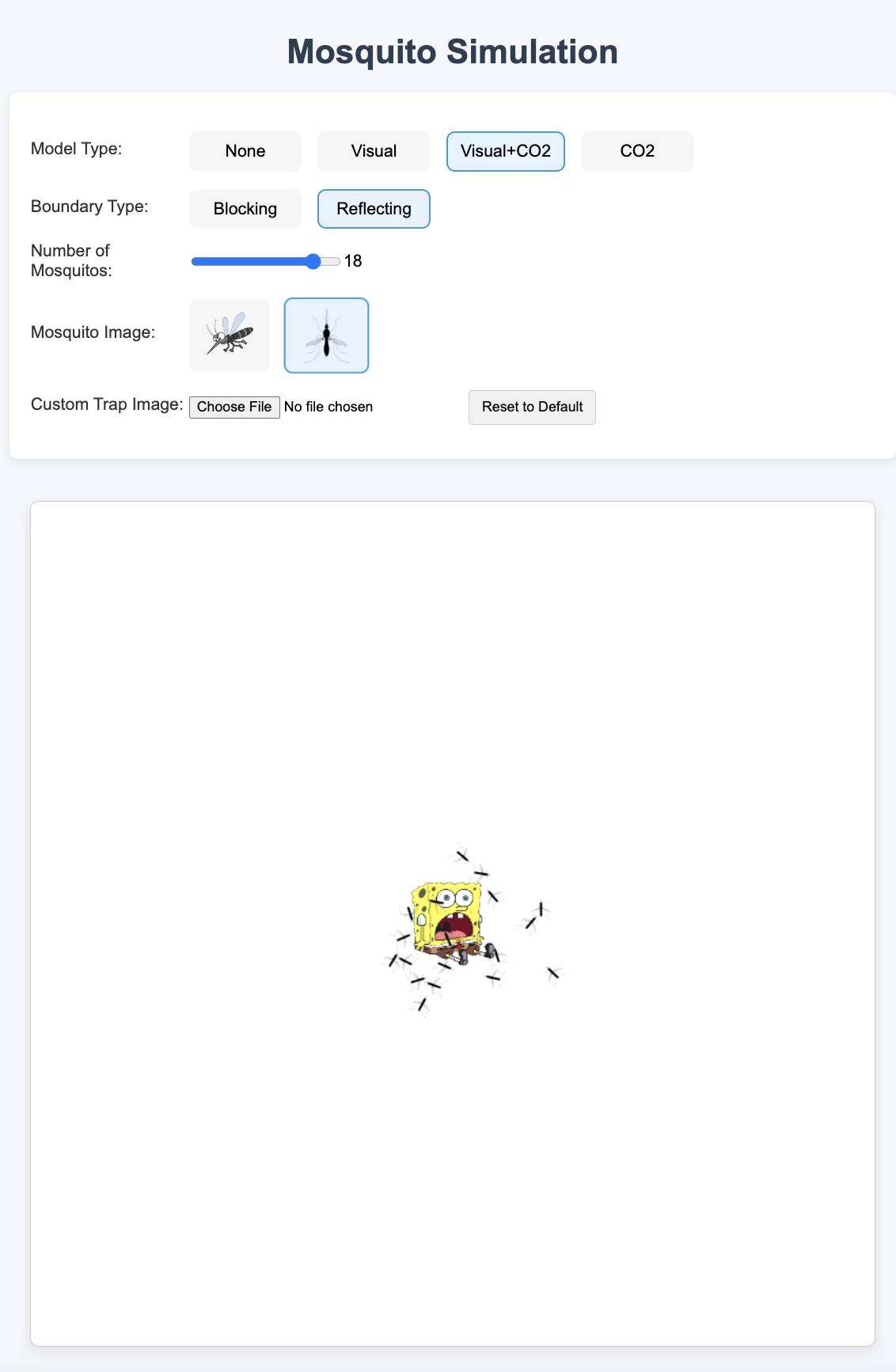}
    \captionsetup{labelsep = period, justification=justified,labelfont=bf,singlelinecheck = false}
    \caption[]
    {
    Screenshot of the interactive mosquito web application.
  }
    \label{figS:website}
\end{figure*}

\begin{figure*}[!t]
    \centering
    \includegraphics[width = 0.9\textwidth]{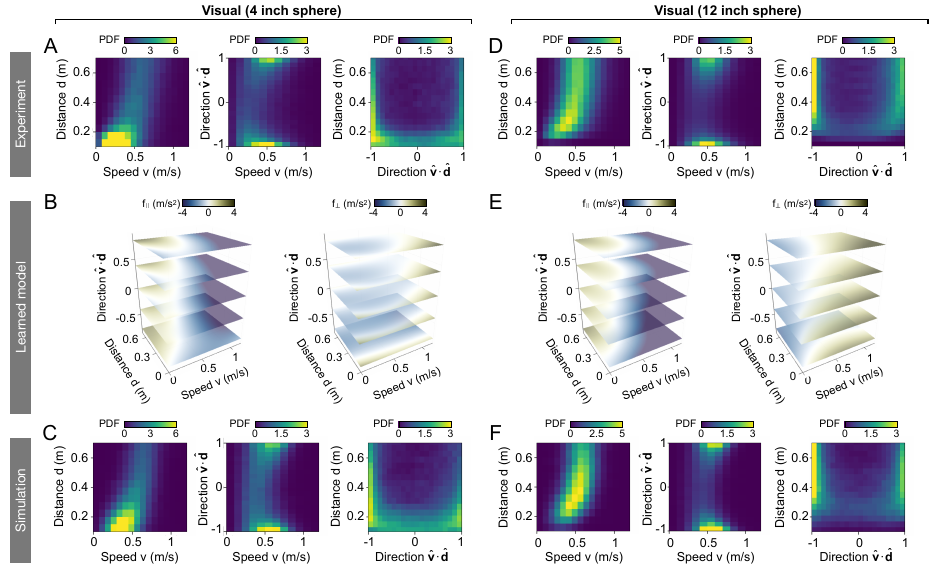}
    \captionsetup{labelsep = period, justification=justified,labelfont=bf,singlelinecheck = false}
    \caption[Mosquito responses to visual targets of varying sizes.]
    {
    {Mosquito responses to visual targets of varying sizes.} 
    We apply the Bayesian inference framework to the ({\bf A}, {\bf D})experimental data to learn ({\bf B}, {\bf E})a dynamical model (see Sec.~\ref{sec:models}), and simulate the model to generate ({\bf C}, {\bf F}) synthetic trajectories. Results are shown for mosquito responses to black spheres with diameters of ({\bf A}--{\bf C}) 4 inches and ({\bf D}--{\bf F}) 12 inches.
    Figure panels correspond to main Fig.~3{\bf C}--{\bf E}.
  }
    \label{figS:visualvarsizes}
\end{figure*}

\begin{figure*}[!t]
    \centering
    \includegraphics[width = .8\textwidth]{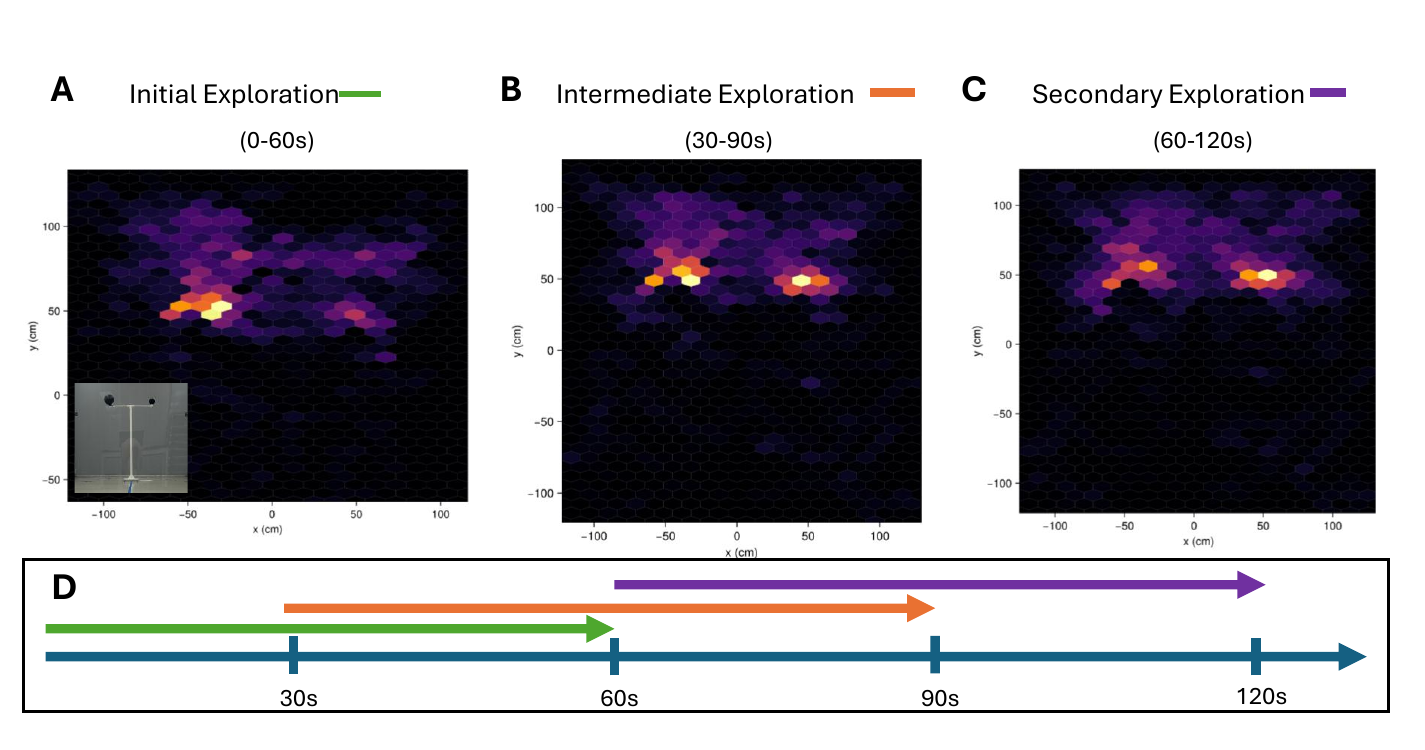}
    \captionsetup{labelsep = period, justification=justified,labelfont=bf,singlelinecheck = false}
    \caption[]
    {
    {Timeline of mosquito trajectory distribution when exposed to two spheres} 
    ({\bf A})  The heat map shows the initial concentration around the larger sphere. Inset shows the setup of experiment where the larger of the two black spheres (12in) is on the left and the smaller (4in) is on the right.
    ({\bf B}) Mosquito heat map moving towards the smaller sphere
    ({\bf C}) Mosquito heat map concentrated around the smaller sphere
    ({\bf D}) Timeline of mosquito trajectory concentrations
  }
    \label{figS:2sphere}
\end{figure*}

To enable readers to interact with our learned mosquito behavioral models, we built an interactive web application.
The application contains the learned models for no cue, visual cue, CO$_2$ cue, and the combined visual + CO$_2$ cues (\SFig\ref{figS:website}).
For visualization purposes on browser screens, the models are converted to 2D.
The cue location in the model is defined at the center of image, which by default is an image of Spongebob but this can be changed through a user upload.
The user can select which model is active in addition to one of two boundary conditions (blocking or reflecting).
The number of mosquitoes can also be changed by the user.
The user can also move the location of the cue (the image) around and watch the mosquitoes respond.

The app works by running a Python server using Flask and a Julia server using Oxygen. 
Information about the user's interaction with the front end is sent to the Flask server, which updates the logic of the boundary conditions and mosquito numbers.
The model is running in the Julia server, so the Flask server sends the positions and velocity data of the mosquitoes to the Julia server, which then provides the time derivatives of these quantities.
Using the derivatives, the Flask server updates the positions and velocities which is then used to update the front end visualization. 
The app is hosted on Railway, which deploys it from a Docker file in a linked GitHub repository.

\clearpage

% SI tables and figures

\begin{table*}[!t]
    \centering
    \includegraphics[width = .8\textwidth]{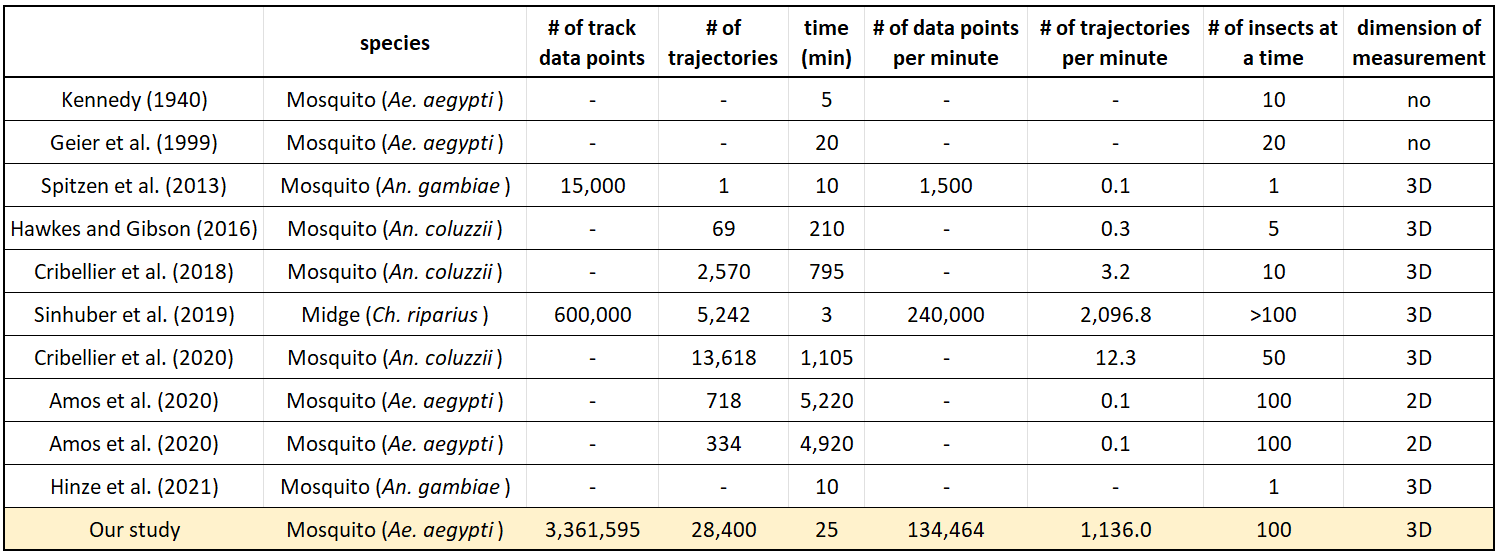}
    \captionsetup{labelsep = period, justification=justified,labelfont=bf,singlelinecheck = false}
    \caption[]
    {
    {Table of experimental metrics comparing previous studies mapping insect trajectories in 3D. Our study values are derived from the "Black 4 in Sphere" experiment.} 
    
  }
    \label{tabS:trajcompare}
\end{table*}

\begin{table*}[!t]
    \centering
    \includegraphics[width = .8\textwidth]{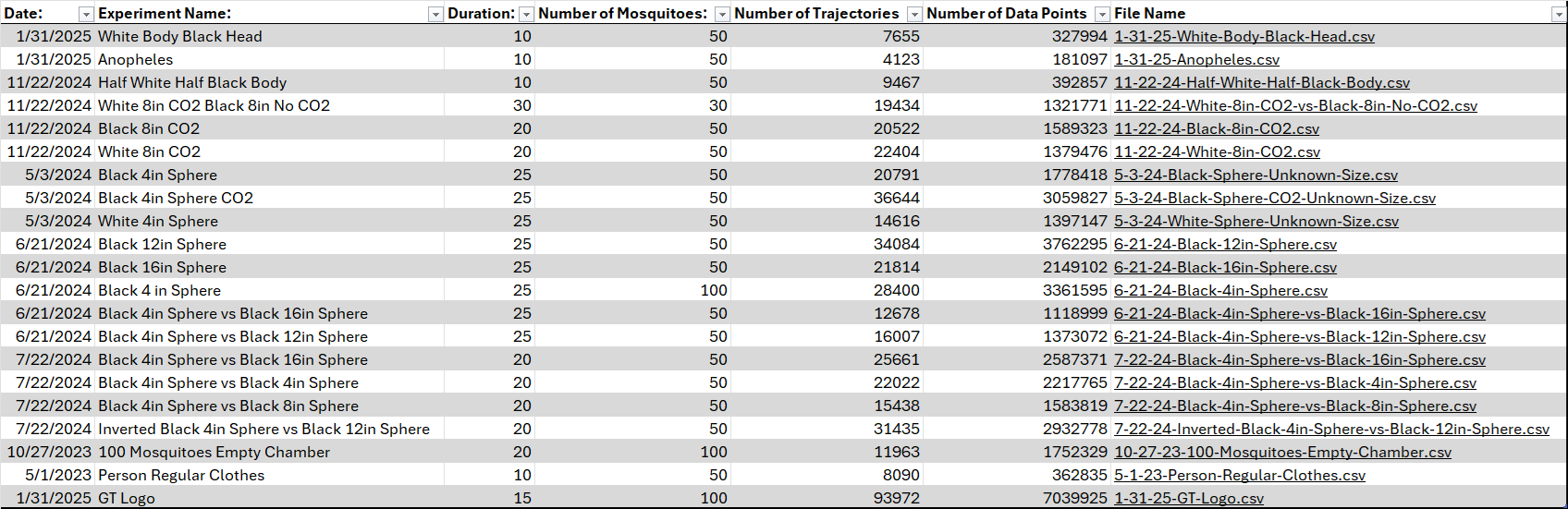}
    \captionsetup{labelsep = period, justification=justified,labelfont=bf,singlelinecheck = false}
    \caption[]
    {
    {Table of trajectories and data-points in each experimental trial} 
  }
    \label{tabS:trajtotal}
\end{table*}

\clearpage

\clearpage

\clearpage

\clearpage

\clearpage

\clearpage

\clearpage

\clearpage

\clearpage

\end{document}